\documentclass[11pt,prd,aps,nofootinbib]{revtex4}

\usepackage{graphicx}
\usepackage{amssymb}
\usepackage{epstopdf}

\newcommand{\be}{\begin{equation}}
\newcommand{\ee}{\end{equation}}
\newcommand{\bea}{\begin{eqnarray}}
\newcommand{\eea}{\end{eqnarray}}

\newcommand{\Romatre}{Dip.~di Matematica e Fisica, Universit\`a  Roma Tre and INFN, Sezione di Roma Tre,\\ Via della Vasca Navale 84, I-00146 Rome, Italy}
\newcommand{\LaSapienza}{Dip.~di Fisica, Universit\`a Roma ``La Sapienza'' and INFN Sezione di Roma,\\ Piazzale Aldo Moro 5, 00185 Roma, Italy}
\newcommand{\RomatreINFN}{Istituto Nazionale di Fisica Nucleare, Sezione di Roma Tre,\\ Via della Vasca Navale 84, I-00146 Rome, Italy}

\begin{document}


\title{Strange and charm HVP contributions to the muon ($g - 2)$\\[2mm] including QED corrections with twisted-mass fermions\vspace{1cm}}

\author{D.~Giusti} \affiliation{\Romatre}
\author{V.~Lubicz} \affiliation{\Romatre}
\author{G.~Martinelli} \affiliation{\LaSapienza}
\author{F.~Sanfilippo} \affiliation{\RomatreINFN}
\author{S.~Simula} \affiliation{\RomatreINFN}

\author{\includegraphics[draft=false]{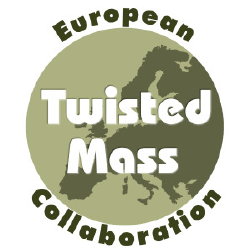}} \noaffiliation


\begin{abstract}
We present a lattice calculation of the Hadronic Vacuum Polarization (HVP) contribution of the strange and charm quarks to the anomalous magnetic moment of the muon including leading-order electromagnetic corrections. 
We employ the gauge configurations generated by the European Twisted Mass Collaboration (ETMC) with $N_f = 2+1+1$ dynamical quarks at three values of the lattice spacing ($a \simeq 0.062, 0.082, 0.089$ fm) with pion masses in the range $M_\pi \simeq 210 - 450$ MeV. 
The strange and charm quark masses are tuned at their physical values. 
Neglecting disconnected diagrams and after the extrapolations to the physical pion mass and to the continuum limit we obtain: $a_\mu^s(\alpha_{em}^2) = (53.1 \pm 2.5) \cdot 10^{-10}$, $a_\mu^s(\alpha_{em}^3) = (-0.018 \pm 0.011) \cdot 10^{-10}$ and $a_\mu^c(\alpha_{em}^2) = (14.75 \pm 0.56) \cdot 10^{-10}$, $a_\mu^c(\alpha_{em}^3) = (-0.030 \pm 0.013) \cdot 10^{-10}$ for the strange and charm contributions, respectively. 
\end{abstract}

\maketitle

\newpage

\section{Introduction}
The  anomalous magnetic moment of the muon $a_\mu \equiv (g -2 ) / 2$ is known experimentally with an accuracy of the order of 0.54 ppm~\cite{Bennett:2006fi}, while the current precision of the Standard Model (SM) prediction is at the level of 0.4 ppm~\cite{PDG}.
The tension of the experimental value with the SM prediction, $a_\mu^{exp} - a_\mu^{SM} = (28.8 \pm 8.0) \cdot 10^{-10}$ \cite{PDG}, corresponds to $\simeq 3.5$ standard deviations and might be an exciting indication of new physics.

The forthcoming $g - 2$ experiments at Fermilab (E989)~\cite{Logashenko:2015xab} and J-PARC (E34)~\cite{Otani:2015lra} aim at reducing the experimental uncertainty by a factor of four, down to 0.14 ppm.
Such a precision makes the comparison of the experimental value of $a_\mu$ with theoretical predictions one of the most important tests of the Standard Model in the quest for new physics effects.
 
It is clear that the experimental precision must be matched by a comparable theoretical accuracy.
With a reduced experimental error, the uncertainty of the hadronic corrections will soon become the main limitation of this test of the SM. 
For this reason an intense research program is under way to improve the evaluation of the leading order hadronic contribution to $a_\mu$ due to the Hadronic Vacuum Polarization (HVP) correction to the one-loop diagram, $a_\mu^{had}(\alpha_{em}^2)$, as well as to the next-to-leading-order hadronic ones. 
The latter include the $O(\alpha_{em}^3)$ contribution of diagrams containing HVP insertions and the leading hadronic light-by-light (LBL) term \cite{Jegerlehner:2009ry}.  

The theoretical predictions for the hadronic contributions have been traditionally obtained from experimental data using dispersion relations for relating the HVP function to the experimental cross section data for $e^+ e^-$ annihilation into hadrons \cite{Davier:2010nc,Hagiwara:2011af}. 
An alternative approach was proposed in Refs.~\cite{Lautrup:1971jf,deRafael:1993za,Blum:2002ii}, namely to compute $a_\mu^{had}(\alpha_{em}^2)$ in Euclidean lattice QCD from the correlation function of two electromagnetic currents. 
In this respect an impressive progress in the lattice determinations of $a_\mu^{had}(\alpha_{em}^2)$ has been achieved in the last few years \cite{Boyle:2011hu,DellaMorte:2011aa,Burger:2013jya,Chakraborty:2014mwa,Chakraborty:2015cso,Bali:2015msa,Chakraborty:2015ugp,Blum:2015you,Blum:2016xpd,Chakraborty:2016mwy,DellaMorte:2017dyu} and very interesting attempts to compute also the LBL contribution are under way both on the lattice \cite{Green:2015sra,Blum:2015gfa} and via dispersion approaches and Chiral Perturbation Theory (ChPT) \cite{Colangelo:2015ama,Bijnens:2016hgx,Colangelo:2017fiz}. 

With the increasing precision of the lattice calculations, it becomes necessary to include electromagnetic (e.m.) and strong isospin breaking (IB) corrections (which contribute at order $O(\alpha_{em}^3)$ and $O(\alpha_{em}^2 (m_d - m_u))$, respectively) to the  HVP.  
In this paper we present the results of a lattice calculation of the leading radiative e.m.~corrections to the HVP contribution due to strange and charm quark intermediate states, obtained using the expansion method of Refs.~\cite{deDivitiis:2011eh,deDivitiis:2013xla}. 
Given the large statistical fluctuations, we are not in the position of giving results for the e.m.~and IB corrections to the HVP contribution from the light up and down quarks, although we will give some details of our computation.  
For the same reason we do not have yet results for the disconnected contributions.

The main results of the present study for $a_\mu^{had}$ are for the lowest-order contributions
 \bea
        a_\mu^s \equiv a_\mu^s(\alpha_{em}^2) & = & (53.1 \pm 2.5) \cdot 10^{-10} ~ , \\
        a_\mu^c \equiv a_\mu^c(\alpha_{em}^2) & = & (14.75 \pm 0.56) \cdot 10^{-10} 
 \eea
 and for the e.m.~corrections
 \bea
        \label{eq:deltamus_final}
        \delta a_\mu^s \equiv a_\mu^s(\alpha_{em}^3) & = & (-0.018 \pm 0.011) \cdot 10^{-10} ~ , \\
        \label{eq:deltamuc_final}
        \delta a_\mu^c \equiv a_\mu^c(\alpha_{em}^3) & = & (-0.030 \pm 0.013) \cdot 10^{-10} ~ .      
 \eea
Our findings demonstrate that the expansion method of Refs.~\cite{deDivitiis:2011eh,deDivitiis:2013xla}, which has been already applied successfully to the calculation of e.m.~and strong IB corrections to meson masses~\cite{deDivitiis:2013xla,Giusti:2017dmp} and leptonic decays of pions and kaons \cite{Lubicz:2016mpj,Tantalo:2016vxk}, works as well also in the case of the HVP contribution to $a_\mu$.
This is reassuring about the feasibility of the determination of the leading e.m.~and strong IB corrections to the HVP contribution from the up and down quarks, which is expected to be not negligible~\cite{Jegerlehner:2009ry} and will be addressed in a separate work.
For a recent calculation of these corrections, though at a large pion mass and at a fixed lattice spacing, see Ref.~\cite{Boyle:2017gzv}, where, as expected, the strong IB effect is found to be at the percent level.
In the strange and charm sectors  the e.m.~corrections (\ref{eq:deltamus_final}-\ref{eq:deltamuc_final}) are found to be negligible with respect to present uncertainties.
 
The paper is organized as follows. 
In section~\ref{sec:master} we introduce the basic quantities and notation. 
In section~\ref{sec:LQCD} we  describe the lattice calculation and give the simulation details.  
In section~\ref{sec:s&c} we present the calculation of the strange and charm contributions to the HPV at order ${\cal{O}}(\alpha_{em}^2)$ and in section~\ref{sec:deltas&c} the corresponding e.m.~corrections at order ${\cal{O}}(\alpha_{em}^3)$, which represent the original part of this work.
Finally, section~\ref{sec:conclusions} contains our conclusions and outlooks for future developments.

\section{Master formula}
\label{sec:master}

The hadronic contribution $a_\mu^{had}$ to the muon anomalous magnetic moment at order $\alpha_{em}^2$ can be related to the Euclidean space-time HVP function $\Pi(Q^2)$ by~\cite{Lautrup:1971jf,deRafael:1993za,Blum:2002ii}
 \be
      a_\mu^{had} = 4 \alpha_{em}^2 \int_0^\infty dQ^2 f(Q^2) \left[ \Pi(Q^2) -  \Pi(0) \right] ~ ,
      \label{eq:amu}
 \ee
where $Q$ is the Euclidean four-momentum and the kinematical kernel $f(Q^2)$ is given by
\be
     f(Q^2) = \frac{1}{m_\mu^2} ~ \frac{1}{\omega} ~ \frac{1}{\sqrt{4 + \omega^2}} ~ \left( \frac{\sqrt{4 + \omega^2} - \omega}{\sqrt{4 + \omega^2} + \omega} \right)^2
     \label{eq:fQ2}
 \ee
with $m_\mu$ being the muon mass and $\omega \equiv Q / m_\mu$.

The HVP form factor $\Pi(Q^2)$ is defined through the HVP tensor as
 \be
      \Pi_{\mu \nu}(Q) \equiv \int d^4x ~ e^{iQ \cdot x} \langle J_\mu(x) J_\nu(0) \rangle = ( \delta_{\mu \nu} Q^2 - Q_\mu Q_\nu ) \Pi(Q^2)
      \label{eq:HVPtensor}
 \ee
where $\langle ... \rangle$ means the average of the $T$-product of the two electromagnetic (e.m.) currents over gluon and fermion fields and 
 \be
     J_\mu(x) \equiv \sum_{f = u, d, s, c, ...} q_f ~ \overline{\psi}_f(x) \gamma_\mu \psi_f(x) 
     \label{eq:Jmu}
 \ee 
with $q_f$ being the electric charge of the quark with flavor $f$ in units of $e$.

In Eq.~(\ref{eq:amu}) the {\it subtracted} HVP function $\Pi_R(Q^2) \equiv \Pi(Q^2) - \Pi(0)$ appears.
This is due to the fact that the photon wave function renormalization constant absorbs the value of the photon self energy at $Q^2 = 0$ in order to guarantee that the e.m.~coupling $\alpha_{em}$ is the experimental one in the limit  $Q^2 \to 0$.

The HVP function $\Pi_R(Q^2)$ can be determined from the vector current-current Euclidean correlator $V(t)$ defined as
 \be
     V(t) \equiv \frac{1}{3} \sum_{i=1,2,3} \int d\vec{x} ~ \langle J_i(\vec{x}, t) J_i(0) \rangle ~ .
     \label{eq:VV}
 \ee
Taking into account that $V(-t) = V(t)$ and choosing $Q$ along the time direction only, one has \cite{Bernecker:2011gh}
 \be
      \Pi_R(Q^2) \equiv \Pi(Q^2) -  \Pi(0) = 2 \int_0^\infty dt ~ V(t) \left[ \frac{\mbox{cos}(Qt) -1}{Q^2} + \frac{1}{2} t^2 \right] ~ .
      \label{eq:cos}
 \ee
Consequently the HVP contribution $a_\mu^{had}$ can be written as
 \be
      a_\mu^{had} = 4 \alpha_{em}^2 \int_0^\infty dt ~ \tilde{f}(t) V(t) ~ ,
      \label{eq:amu_t}
 \ee
 where $\tilde{f}(t)$ is given by \cite{Bernecker:2011gh}
  \be
      \tilde{f}(t) \equiv 2 \int_0^\infty dQ^2 ~ f(Q^2) \left[ \frac{\mbox{cos}(Qt) -1}{Q^2} + \frac{1}{2} t^2 \right] ~ .
      \label{eq:ftilde}
  \ee

In what follows we will limit ourselves to the connected contributions to $a_\mu^{had}$.
In this case each quark flavor $f$ contributes separately, i.e.
 \be
     a_\mu^{had} \sim \sum_{f = u, d, s, c, ...} [a_\mu^{had}(f)]_{(conn)} ~ .
     \label{eq:amuf}
 \ee
For sake of simplicity we drop the label $f$ and the suffix $(conn)$, but it is understood that hereafter we refer to the connected part of $a_\mu^{had}$ for a generic quark flavor $f$.

\section{Lattice QCD simulations for $a_\mu^{had}$}
\label{sec:LQCD}

The vector correlator $V(t)$ can be calculated on a lattice with volume $L^3$ and temporal extension $T$ at discretized values of $\overline{t} \equiv t / a$ from $-\overline{T}/2$ to $\overline{T}/2$ with $\overline{T} = T / a$. 
From now on all the ``overlined'' quantities are in lattice units.

A natural procedure is to split Eq.~(\ref{eq:amu_t}) into two contributions, $a_\mu^{had}(<)$ and $a_\mu^{had}(>)$, corresponding to $0 \leq \overline{t} \leq \overline{T}_{data}$ and $\overline{t} > \overline{T}_{data}$, respectively.
In the first contribution $a_\mu^{had}(<)$ the vector correlator is directly given by the lattice data, while for the second contribution $a_\mu^{had}(>)$ an analytic representation is required (see Refs.~\cite{Chakraborty:2015ugp,Blum:2015you,Chakraborty:2016mwy,DellaMorte:2017dyu}).
If $\overline{T}_{data}$ is large enough that the ground-state contribution is dominant for $\overline{t} > \overline{T}_{data}$, one can write
\be
     a_\mu^{had} = a_\mu^{had}(<) + a_\mu^{had}(>) 
     \label{eq:decomposition}
 \ee
with
 \bea 
     \label{eq:amu_cuts<}
     a_\mu^{had}(<) & = & 4 \alpha_{em}^2 \sum_{\overline{t} = 0}^{\overline{T}_{data}} \overline{f}(\overline{t}) \overline{V}(\overline{t}) ~ , \\
    \label{eq:amu_cuts>}
     a_\mu^{had}(>) & = & 4 \alpha_{em}^2 \sum_{\overline{t} = \overline{T}_{data} + 1}^\infty \overline{f}(\overline{t}) 
                                        \frac{\overline{Z}_V}{2 \overline{M}_V} e^{- \overline{M}_V \overline{t}} ~ ,
 \eea
where $\overline{Z}_V \equiv (1/3) \sum_{i=1,2,3} | \langle 0| J_i(0) | V \rangle |^2$ is the (squared) matrix element of the vector current operator (for the given quark flavor $f$) between the vector ground-state and the vacuum. 

Note that $\overline{T}_{data}$ cannot be taken equal to $\overline{T} / 2$, because on the lattice the vector correlator possesses backward signals. 
In order to avoid them one has to choose an upper limit $\overline{T}_{data}$ sufficiently smaller than $\overline{T} / 2$.
An important consistency check is that the sum of the two terms in the r.h.s.~of Eq.~(\ref{eq:decomposition}) should be almost independent of the specific choice of the value of $\overline{T}_{data}$, as it will be shown later in Section \ref{sec:s&c}.

\subsection{Simulation details}
\label{sec:simulations}

The gauge ensembles used in this work are the same adopted in Ref.~\cite{Carrasco:2014cwa} to determine the up, down, strange and charm quark masses. 
We employed the Iwasaki action \cite{Iwasaki:1985we} for gluons  and the Wilson Twisted Mass Action \cite{Frezzotti:2000nk, Frezzotti:2003xj, Frezzotti:2003ni} for sea quarks . 
In order to avoid the mixing of strange and charm quarks in the valence sector we adopted a non-unitary set up \cite{Frezzotti:2004wz} in which the valence strange and charm quarks are regularized as Osterwalder-Seiler fermions \cite{Osterwalder:1977pc}, while the valence up and down quarks have the same action of the sea.
Working at maximal twist such a setup guarantees an automatic ${\cal{O}}(a)$-improvement \cite{Frezzotti:2003ni, Frezzotti:2004wz}.

We considered three values of the inverse bare lattice coupling $\beta$ and different lattice volumes, as shown in Table \ref{tab:simudetails}, where the number of configurations analyzed ($N_{cfg}$) corresponds to a separation of $20$ trajectories.
At each lattice spacing, different values of the light sea quark masses have been considered. 
The light valence and sea quark masses are always taken to be degenerate. 
The bare masses of both the strange ($a\mu_s$) and the charm ($a\mu_c$) valence quarks are obtained, at each $\beta$, using the physical strange and charm masses and the mass renormalization constant (RC) determined in Ref.~\cite{Carrasco:2014cwa}.
The values of the lattice spacing are: $a = 0.0885(36)$, $0.0815(30)$, $0.0619(18)$ fm at $\beta = 1.90$, $1.95$ and $2.10$, respectively.

\begin{table}[hbt!]
{\footnotesize
\begin{center}
\renewcommand{\arraystretch}{1.20}
\begin{tabular}{||c|c|c||c|c|c|c|c|c||c|c|c||}
\hline
ensemble & $\beta$ & $V / a^4$ &$a\mu_\ell$&$a\mu_\sigma$&$a\mu_\delta$&$N_{cfg}$& $a\mu_s$ & $a\mu_c$ & $M_\pi$ & 
$M_K$ & $M_D$ \\
\hline \hline
$A30.32$ & $1.90$ & $32^{3}\times 64$ &$0.0030$ &$0.15$ &$0.19$ &$150$&  $0.02363$ & $0.27903$ & 275 (10) & 568 (22) & 2012 (77) \\
$A40.32$ & & & $0.0040$ & & & $100$ & & & 316 (12) & 578 (22) & 2008 (77) \\
$A50.32$ & & & $0.0050$ & & &  $150$ & & & 350 (13) & 586 (22) & 2014 (77) \\
\cline{1-1} \cline{3-4}
$A40.24$ & & $24^{3}\times 48 $ & $0.0040$ & & & $150$ & & & 322 (13) & 582 (23) & 2017 (77) \\
$A60.24$ & & & $0.0060$ & & &  $150$ & & & 386 (15) & 599 (23) & 2018 (77) \\
$A80.24$ & & & $0.0080$ & & &  $150$ & & & 442 (17) & 618 (24) & 2032 (78) \\
$A100.24$ &  & & $0.0100$ & & &  $150$ & & & 495 (19) & 639 (24) & 2044 (78) \\
\cline{1-1} \cline{3-4}
$A40.20$ & & $20^{3}\times 48 $ & $0.0040$ & & & $150$ & & & 330 (13) & 586 (23) & 2029 (79) \\
\hline \hline
$B25.32$ & $1.95$ & $32^{3}\times 64$ &$0.0025$&$0.135$ &$0.170$& $150$& $0.02094$ & $0.24725$ & 259 ~(9) & 546 (19) & 1942 (67) \\
$B35.32$ & & & $0.0035$ & & & $150$ & & & 302 (10) & 555 (19) & 1945 (67) \\
$B55.32$ & & & $0.0055$ & & & $150$ & & & 375 (13) & 578 (20) & 1957 (68) \\
$B75.32$ &  & & $0.0075$ & & & $~80$ & & & 436 (15) & 599 (21) & 1970 (68) \\
\cline{1-1} \cline{3-4}
$B85.24$ & & $24^{3}\times 48 $ & $0.0085$ & & & $150$ & & & 468 (16) & 613 (21) & 1972 (68) \\
\hline \hline
$D15.48$ & $2.10$ & $48^{3}\times 96$ &$0.0015$&$0.1200$ &$0.1385 $& $100$& $0.01612$ & $0.19037$ & 223 ~(6) & 529 (14) & 1929 (49) \\ 
$D20.48$ & & & $0.0020$ & & & $100$ & & & 256 ~(7) & 535 (14) & 1933 (50) \\
$D30.48$ & & & $0.0030$ & & & $100$ & & & 312 ~(8) & 550 (14) & 1937 (49) \\
 \hline   
\end{tabular}
\renewcommand{\arraystretch}{1.0}
\end{center}
}
\vspace{-0.25cm}
\caption{\it \small Values of the simulated quark bare masses (in lattice units), of the pion, kaon and $D$-meson masses (in units of MeV) for the $16$ ETMC gauge ensembles with $N_f = 2+1+1$ dynamical quarks used in this work (see Ref.~\cite{Carrasco:2014cwa}). The values of the strange and charm quark bare masses $a \mu_s$ and $a \mu_c$, given for each gauge ensemble, correspond to the physical strange and charm quark masses determined in Ref.~\cite{Carrasco:2014cwa}. The central values and errors of the pion, kaon and $D$-meson masses are evaluated using the bootstrap events of the eight branches of the analysis of Ref.~\cite{Carrasco:2014cwa}.}
\label{tab:simudetails}
\end{table}

In this work we made use of the bootstrap samplings elaborated for the input parameters of the quark mass analysis of Ref.~\cite{Carrasco:2014cwa}.
There, eight branches of the analysis were adopted differing in: 
\begin{itemize}
\item the continuum extrapolation adopting for the scale parameter either the Sommer parameter $r_0$ or the mass of a fictitious PS meson made up of strange(charm)-like quarks; 
\item the chiral extrapolation performed with fitting functions chosen to be either a polynomial expansion or a Chiral Perturbation Theory (ChPT) Ansatz in the light-quark mass;
\item the choice between the methods M1 and M2, which differ by $O(a^2)$ effects, used to determine in the RI'-MOM scheme the mass RC $Z_m = 1 / Z_P$. 
\end{itemize}

\begin{figure}[htb!]
\centering{\scalebox{0.80}{\includegraphics{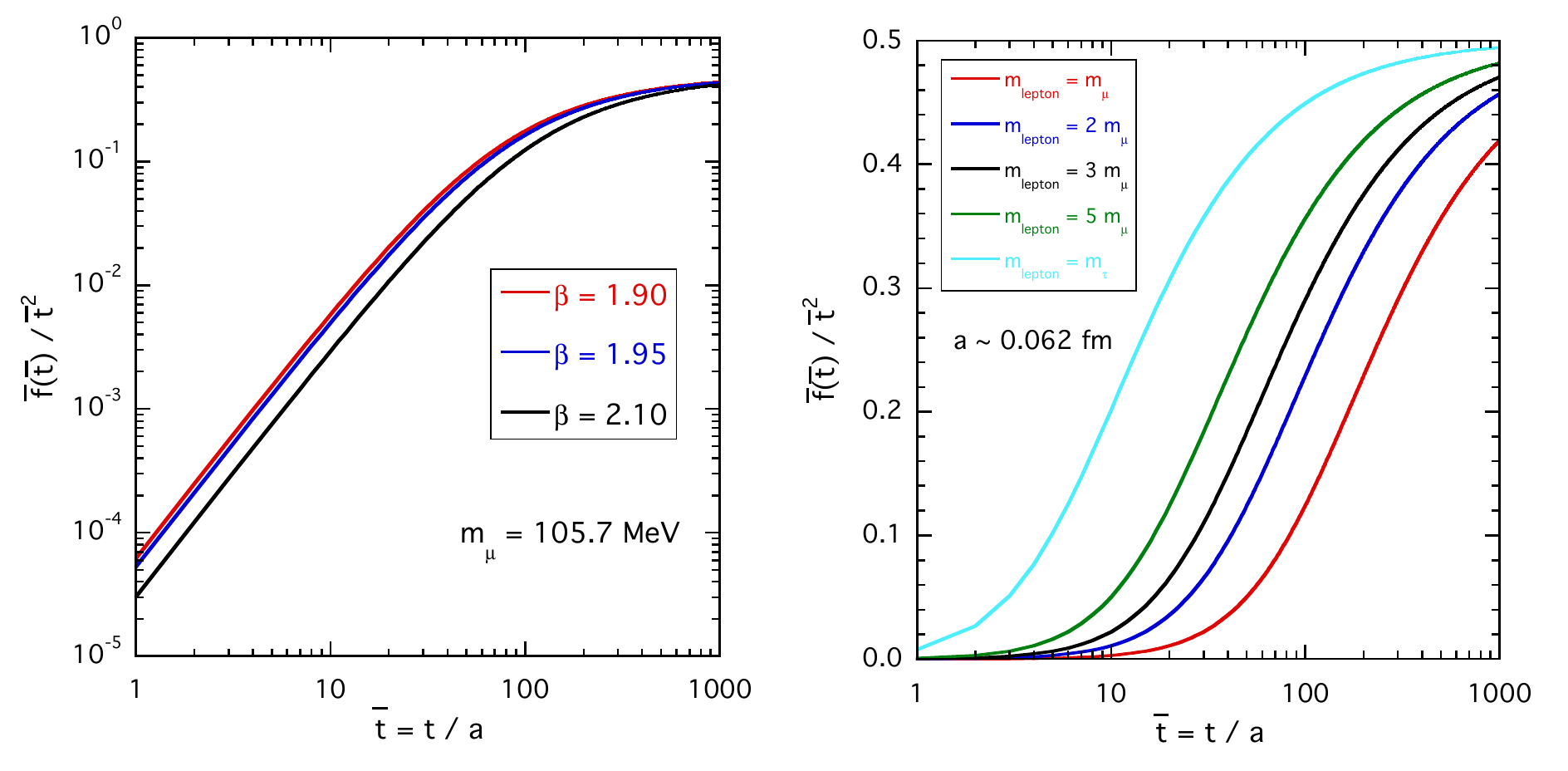}}}
\caption{\it \small The kernel function $\overline{f}(\overline{t}) / \overline{t}^2$ versus the time distance $\overline{t} = t / a$ evaluated for the three values of the lattice spacing corresponding to the ETMC ensembles of Table \ref{tab:simudetails} (left panel) and for various values of the lepton mass ranging from the $\mu$ to the $\tau$ mass (right panel).}
\label{fig:Ftau}
\end{figure}

The kernel function $\overline{f}(\overline{t})$, appearing in Eqs.~(\ref{eq:amu_cuts<}-\ref{eq:amu_cuts>}), is explicitly given by
 \be
      \overline{f}(\overline{t}) = \frac{4}{\overline{m}_\mu^2} \int_0^\infty d\omega ~ \frac{1}{\sqrt{4 + \omega^2}} ~ 
                                              \left( \frac{\sqrt{4 + \omega^2} - \omega}{\sqrt{4 + \omega^2} + \omega} \right)^2
                                              \left[ \frac{\mbox{cos}(\omega \overline{m}_\mu \overline{t}) - 1}
                                              {\omega^2} + \frac{1}{2} \overline{m}_\mu^2 \overline{t}^2 \right] 
      \label{eq:ftau}
 \ee
and can be easily calculated numerically at any value of $\overline{t}$ for given values of the muon mass in lattice units, $\overline{m}_\mu \equiv a m_\mu$.
This is shown in Fig.~\ref{fig:Ftau} in the case of the muon at the three values of the lattice spacing of the ETMC ensembles of Table \ref{tab:simudetails} (left panel) and for various values of the lepton mass (right panel) ranging from the $\mu$ to the $\tau$ mass.
The kernel function $\overline{f}(\overline{t})$ is proportional to $\overline{t}^4$ at small values of $\overline{t}$, diverges as $\overline{t}^2$ at large values of the time distance and has some sensitivity to the value of the lattice spacing. 
Instead it changes significantly with the mass of the lepton enhancing the role of the large-time behaviour of the vector correlator in the case of light leptons.

\subsection{Local versus conserved vector currents on the lattice}
\label{sec:currents}

The vector correlator $\overline{V}(\overline{t})$ can be calculated using either the lattice conserved vector current $J_\mu^C(x)$ or the local one $J_\mu(x)$. 
The latter needs to be renormalized and in our twisted-mass setup the local vector current for each quark flavor $f$ is given by
 \be
     J_\mu(x) = q_f ~ Z_V ~ \bar{\psi}_f(x) \gamma_\mu \psi_f(x) ~ ,
     \label{eq:localVV}
 \ee
where, being at maximal twist, the renormalization is multiplicative through the renormalization constant $Z_V$.

The variation of the lattice action with respect to a vector rotation $\alpha_V(x)$ of the quark fields, i.e.~$\psi(x) \to e^{i q_f \alpha_V(x)} ~ \psi(x)$ and $\overline{\psi}(x) \to \overline{\psi}(x) ~ e^{-i q_f \alpha_V(x)}$ (for any quark flavor $f$), provides the relevant Ward-Takahashi identity for the conserved current $J_\mu^C$ expressed in terms of the backward lattice derivative.
In our twisted-mass setup one has 
 \bea
     J_\mu^C(x) & = & q_f ~ \frac{1}{2} \left[ \bar{\psi}_f(x) (\gamma_\mu - i \tau^3 \gamma_5 ) U_\mu(x) \psi_f(x + a \hat{\mu}) \right.
                                  \nonumber \\
                        & + & \left. \bar{\psi}_f(x + a \hat{\mu}) (\gamma_\mu + i \tau^3 \gamma_5 ) U_\mu^\dagger(x) \psi_f(x) \right] ~ .
     \label{eq:conservedV}
 \eea
According to the vector Ward-Takahashi identity the polarization tensor $\langle J_\mu^C(x) J_\nu^C(y) \rangle$ is not transverse because of the contact term arising from the vector rotation of the conserved current $J_\nu^C(y)$, which generates the backward lattice derivative of the tadpole operator and is power divergent as $1 / a^3$.
Thus, in the case of two conserved currents the transverse HVP tensor is defined as
 \be
     \Pi_{\mu \nu}^{CC}(x, y) \equiv \langle J_\mu^C(x) J_\nu^C(y) \rangle - \frac{1}{a^3} \delta_{\mu \nu} \delta_{x y}  \langle T_\nu(y) \rangle ~ ,
     \label{eq:CC}
 \ee
where the tadpole operator is explicitly given by
 \bea
     T_\nu(y) & = & q_f^2 ~ \frac{1}{2} \left[ \bar{\psi}_f(y) (\gamma_\nu - i \tau^3 \gamma_5 ) U_\nu(y) \psi_f(y+ a \hat{\nu}) \right. 
                             \nonumber \\
                    & - & \left. \bar{\psi}_f(y + a \hat{\nu}) (\gamma_\nu + i \tau^3 \gamma_5 ) U_\nu^\dagger(y) \psi_f(y) \right] ~ .
     \label{eq:tadpole}
 \eea
On the contrary, in the case of one conserved and one local currents there is no contact term because the vector rotation of the local current (\ref{eq:localVV}) is zero.
One gets
 \be
     \Pi_{\mu \nu}^{CL}(x, y) \equiv \langle J_\mu^C(x) J_\nu(y) \rangle ~ ,
     \label{eq:CL}
 \ee
which is transverse only with respect to the $\mu$ index (i.e., $\partial_\mu^b  \Pi_{\mu \nu}^{CL}(x, y) = 0$, where $\partial_\mu^b$ is the backward lattice derivative).

In the case of two local currents the polarization tensor $\langle J_\mu(x) J_\nu(y) \rangle$ is not transverse.
The mixing pattern of the product of two local currents with all possible operators with equal and lower dimensions has been investigated for the twisted-mass setup in Ref.~\cite{Burger:2014ada}. 
The outcome is that at maximal twist one has
 \bea
      \int d^4x e^{i Q x} \langle J_\mu(x) J_\nu(0) \rangle & = & \Pi_{\mu \nu}(Q) + \delta_{\mu \nu} {\cal{Z}}_1 
            \left( \frac{1}{a^2} - S_6 + \frac{S_5^2}{2} \right) + \delta_{\mu \nu} {\cal{Z}}_m m^2 \nonumber \\
            & + & \delta_{\mu \nu} {\cal{Z}}_L Q^2 + \left( \delta_{\mu \nu} Q^2 - Q_\mu Q_\nu \right) {\cal{Z}}_T + {\cal{O}}(a^2) ~ ,
      \label{eq:mixing}                                          
 \eea
where $\Pi_{\mu \nu}(Q)$ is the transverse polarization tensor, $S_5$ and $S_6$ are the vacuum expectation values of the dimension-5 and -6 terms of the Symanzik expansion of the twisted-mass action, $m$ is the (twisted) quark mass and the quantities ${\cal{Z}}_1$, ${\cal{Z}}_m$, ${\cal{Z}}_L$ and ${\cal{Z}}_T$ are mixing coefficients.

In the r.h.s.~of Eq.~(\ref{eq:mixing}) the second and third terms do not depend on $Q$, while the fourth and fifth terms are $Q$-dependent.
The former ones can be eliminated by considering the subtracted form
 \bea
      \int d^4x \left( e^{i Q x} - 1 \right) \langle J_\mu(x) J_\nu(0) \rangle & = & \Pi_{\mu \nu}(Q) + \delta_{\mu \nu} {\cal{Z}}_L Q^2 \nonumber \\
                                                                                                               & + & \left( \delta_{\mu \nu} Q^2 - Q_\mu Q_\nu \right) {\cal{Z}}_T + 
                                                                                                                        {\cal{O}}(a^2) ~ ,
      \label{eq:sub1}                                          
 \eea
where we have considered that $\Pi_{\mu \nu}(0) = 0$ in the infinite volume limit \cite{Bernecker:2011gh}.
Choosing $Q$ along the time direction only with $\mu = \nu = i =1, 2, 3$ one has 
 \be
     \int dt \left( e^{i Q t} - 1 \right) \int d^3x  \langle J_i(x) J_i(0) \rangle = \Pi_{ii}(Q) + ({\cal{Z}}_L + {\cal{Z}}_T) Q^2 + {\cal{O}}(a^2) ~ .
 \ee
Using Eqs.~(\ref{eq:HVPtensor}) and (\ref{eq:VV}) and taking into account that $V(t) = V(-t)$, one obtains
 \be
     2 \int_0^\infty dt \frac{\mbox{cos}(Qt) - 1}{Q^2} V(t) = \Pi(Q^2) + {\cal{Z}}_L + {\cal{Z}}_T + {\cal{O}}(a^2) ~ .
 \ee
In the the renormalized HVP function $[\Pi(Q^2) - \Pi(0)]$ the term (${\cal{Z}}_L + {\cal{Z}}_T$) cancels out, so that Eq.~(\ref{eq:cos}) is recovered and the ${\cal{O}}(a)$-improvement of the renormalized HVP function is guaranteed.

In this work the local version of the vector current is adopted (see later Eq.~(\ref{eq:localV}) in Section \ref{sec:GS}).

\subsection{Perturbative QCD (pQCD) and the behavior of $V(t)$ at small $t$}
\label{sec:pQCD}

The HVP function $\Pi_R(Q^2)$ obeys the (once subtracted) dispersion relation
 \be
     \Pi_R(Q^2) \equiv \Pi(Q^2) -  \Pi(0) = \frac{1}{12 \pi^2} \int_{s_{thr}}^{\infty} ds \frac{Q^2}{s (s + Q^2)} R^{had}(s) ~ ,
     \label{eq:HVPdisp}
 \ee
where $R^{had}(s)$ is related to the (one photon) $e^+ e^-$ annihilation cross section into hadrons, $\sigma^{had}(s)$, by
 \be
      \sigma^{had}(s) = \frac{4 \pi \alpha_{em}^2}{s} R^{had}(s) 
      \label{eq:Rhad}
 \ee
with $s$ being the center-of-mass energy and $s_{thr} = 4 M_\pi^2$.

The pQCD prediction for $R^{had}(s)$ is known up to three loops including mass corrections \cite{Chetyrkin:1996cf}.
Here we limit ourselves to the lowest order prediction, which, for each quark flavor $f$, reads as
 \be
    R^{pQCD}(s) = q_f^2 N_c \sqrt{1 - \frac{4 m^2}{s}} \left( 1 + \frac{2m^2}{s} \right) \theta(s - 4 m^2) + {\cal{O}}(\alpha_s) ~ ,
    \label{eq:RpQCD}
 \ee
where $m$ is the on-shell quark mass.
Inserting Eq.~(\ref{eq:RpQCD}) into Eq.~(\ref{eq:HVPdisp}) one obtains
 \be
      \Pi_R^{pQCD}(Q^2) = \frac{q_f^2 N_c}{12 \pi^2} \left[ x^2 - \frac{5}{3} + (2 - x^2) \sqrt{1 + x^2} ~ 
                                          \mbox{ln}\left( \frac{1 + \sqrt{1 + x^2}}{x} \right) \right] ~ ,
      \label{eq:PiRpQCD}
 \ee
where $x \equiv 2 m / Q$.
The behavior of $\Pi_R^{pQCD}(Q^2)$ at large $Q^2$ is given by
 \bea
     \Pi_R^{pQCD}(Q^2) ~ & _{\overrightarrow{Q^2 \to \infty}} & ~ \frac{q_f^2 N_c}{12 \pi^2} \left\{ \mbox{ln}\left( \frac{Q^2}{m^2} \right) -
                                         \frac{5}{3} + 6 ~ \frac{m^2}{Q^2}  \right. \nonumber \\
                                        & & \left. - 3 ~ \frac{m^4}{Q^4} \left[ 1 + 2 \mbox{ln}\left( \frac{Q^2}{m^2} \right) \right] + 
                                        {\cal{O}}\left( \frac{m^6}{Q^6} \right) \right\} ~ ,
     \label{eq:PiRpQCD_largeQ2}
 \eea
which exhibits a logarithmic divergence.

In the continuum the vector correlator (\ref{eq:VV}) can be obtained simply by taking the Fourier transform of the spatial components of the HVP tensor (\ref{eq:HVPtensor}).
Choosing $Q$ along the time direction only, one gets
 \be
     V(t) \equiv \int_{-\infty}^{\infty} dQ ~ e^{-i Q t} \frac{1}{3} \sum_{i = 1, 2, 3} \Pi_{ii}(Q) ~ _{\overrightarrow{t^{\phantom{1}} > 0}} ~ 
          \int_{-\infty}^{\infty} dQ ~ e^{-i Q t} Q^2 \Pi_R(Q^2) ~ .
     \label{eq:Vt_continuum}
 \ee
Using Eq.~(\ref{eq:HVPdisp}) one has
 \bea
     V(t) & ~ _{\overrightarrow{t^{\phantom{1}} > 0}} ~ &  \frac{1}{12 \pi^2} \int_{s_{thr}}^{\infty} ds \frac{1}{s} R^{had}(s) 
               \int_{-\infty}^{\infty} dQ ~ e^{-i Q t}  \frac{Q^4}{s + Q^2} \nonumber \\
           & ~ _{\overrightarrow{t^{\phantom{1}} > 0}} ~ & \frac{1}{24 \pi^2} \int_{s_{thr}}^{\infty} ds \sqrt{s} R^{had}(s) e^{-\sqrt{s} t}  ~ .
      \label{eq:Vt_R}
 \eea
Consequently, using the pQCD result~(\ref{eq:RpQCD}) for $R^{had}(s)$ (including the quark mass threshold $s_{thr} = 4 m^2$) the pQCD prediction for $V(t)$ is given by
 \bea
     V^{pQCD}(t) ~ & _{\overrightarrow{t^{\phantom{1}} > 0}} & ~ \frac{2 q_f^2 N_c}{3 \pi^2} m^3 \int_1^{\infty} dy ~ y^2 \sqrt{1 - \frac{1}{y^2}} 
                                                                              \left( 1 + \frac{1}{2 y^2} \right) e^{- 2m t y} \nonumber \\
                             & = & \frac{q_f^2 N_c}{6 \pi^2} \left\{ \frac{1}{t^3} e^{-2mt}(1 + 2 mt + 2 m^2 t^2) \right. \nonumber \\
                             & + & \left. 4 m^3 \int_1^{\infty} dy ~ y^2 \left[ \sqrt{1- \frac{1}{y^2}} \left( 1 + \frac{1}{2 y^2} \right) -1 \right] 
                                       e^{- 2m t y} \right\} ~ .
     \label{eq:Vt_pQCD}
 \eea
Note that $e^{-2mt}(1 + 2 mt + 2 m^2 t^2) = 1 + {\cal{O}}(m^3 t^3)$ and therefore at small values of $t$ the vector correlator $V^{pQCD}(t)$ is dominated by a mass-independent term, namely
 \be
      V^{pQCD}(t) ~ _{\overrightarrow{t^{\phantom{1}} << 1/m}}  ~ \frac{q_f^2}{2 \pi^2} \frac{1}{t^3} + {\cal{O}}(m^3, m^4t) ~ ,
      \label{eq:Vt_pQCD_smallt}
 \ee
which represents also the vector correlator $V^{pQCD}(t)$ in the massless limit.

\begin{figure}[htb!]
\centering{\scalebox{0.80}{\includegraphics{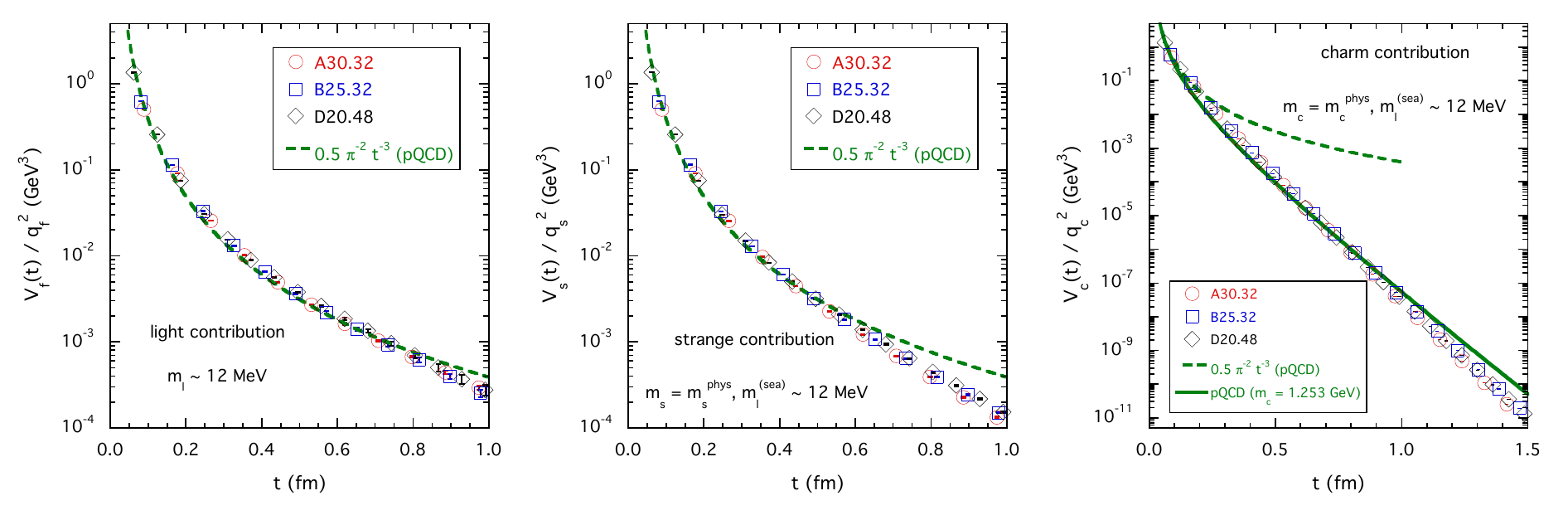}}}
\caption{\it \small The vector correlator $V(t) / q_f^2$ (in physical units) in the case of the light (left panel), strange (middle panel) and charm (right panel) contributions for the ETMC gauge ensembles specified in the inset, which share an approximate common value of the light-quark mass $m_\ell \simeq 12$ MeV and differ in the values of the lattice spacing. The dashed lines represent the mass-indepedent pQCD prediction (\ref{eq:Vt_pQCD_smallt}), while the solid line in the right panel is the pQCD prediction (\ref{eq:Vt_pQCD}) calculated for an on-shell charm quark mass equal to $m_c = 1.253$ GeV.}
\label{fig:VpQCD}
\end{figure}

In Fig.~\ref{fig:VpQCD} we compare the pQCD predictions (\ref{eq:Vt_pQCD}) and (\ref{eq:Vt_pQCD_smallt}) with the vector correlator $V(t)$ obtained using the ETMC ensembles A30.32, B25.32 and D20.48, which share an approximate common value of the light-quark mass $m_\ell \simeq 12$ MeV and differ only in the values of the lattice spacing.
It can be clearly seen that at small values of $t$ the lattice data match nicely the (lowest order) pQCD prediction.
The inclusion of the radiative corrections from Ref.~\cite{Chetyrkin:1996cf} leads to an effect of the order of  $\approx 10 \%$, which does not modify the quality of  the agreement shown in Fig.~\ref{fig:VpQCD}.

A closer look to Fig.~\ref{fig:VpQCD} shows that the matching with pQCD is present up to time distances of $\approx 1$ fm (the agreement can be extended in the case of the strange vector correlator by including the corrections due to the strange quark mass), which corresponds to $1 / \Lambda_{QCD}$ with $\Lambda_{QCD} \approx 300$ MeV, i.e., the agreement is observed down to energy scales of the order of $\Lambda_{QCD}$.
One would expect that pQCD works at $t << 1 / \Lambda_{QCD}$ or $Q >> \Lambda_{QCD}$.
The fact that instead the matching appears to work at larger time distances is a nice manifestation of the quark-hadron duality {\it \`a la SVZ}, which states that the sum of the contributions of the excited states is dual to the pQCD behaviour.

Finally, it is interesting to estimate the contribution to $a_\mu^{had}$ coming from values of $Q^2$ larger than $Q_{max}^2 \simeq 1 / a^2$, which for our lattice setup is always larger than $4$ GeV$^2$.
Using the pQCD prediction (\ref{eq:PiRpQCD}) for the large $Q^2$-behavior of $\Pi_R(Q^2)$, one gets: $a_\mu^{had}(Q^2 > 4~\mbox{GeV}^2) \simeq 1.3, 0.11, 0.06$ (in units of $10^{-10}$) in the case of the light, strange and charm contributions, respectively.
The above findings represent only a small fraction of the uncertainties of the present lattice estimates of the three contributions to $a_\mu^{had}$ (see Refs.~\cite{Burger:2013jya,Chakraborty:2015cso,Chakraborty:2014mwa,Blum:2016xpd}).

Alternatively we can check the change induced in the kernel function $\overline{f}(\overline{t})$ by cutting the upper integration limit in Eq.~(\ref{eq:ftau}) to $\omega_{max} = Q_{max} / m_\mu \simeq 1 / (am_\mu)$.
Since in our lattice setup $\omega_{max} \gtrsim 20$, the kernel function $\overline{f}(\overline{t})$ changes at most by one part over $\simeq 10^6$ at small $\overline{t}$ in the case of the muon.

\subsection{Ground-state identification}
\label{sec:GS}

Our numerical simulations of the vector correlator $V(t)$ have been carried out in the context of a more general project aiming at the determination of the e.m.~and strong IB corrections to pseudoscalar meson masses and leptonic decay constants \cite{PRACE}.
In this project the bilinear operators were constructed adopting opposite values of the Wilson $r$-parameter.
Thus, instead of Eq.~(\ref{eq:localVV}) the evaluation of the vector correlator has been carried out using the following local current:
 \be
    J_\mu(x) = Z_A ~ q_f ~ \bar{\psi}_{f^\prime}(x) \gamma_\mu \psi_f(x) ~ ,
    \label{eq:localV}
 \ee
where $\psi_{f^\prime}$ and $\psi_f$ represent two quarks with the same mass and charge, but regularized with opposite values of the Wilson $r$-parameter, i.e.~$r_{f^\prime} = - r_f$.
Being at maximal twist the current (\ref{eq:localV}) renormalizes multiplicatively with the renormalization constant $Z_A$ of the axial current.

\begin{figure}[htb!]
\centering{\scalebox{0.75}{\includegraphics{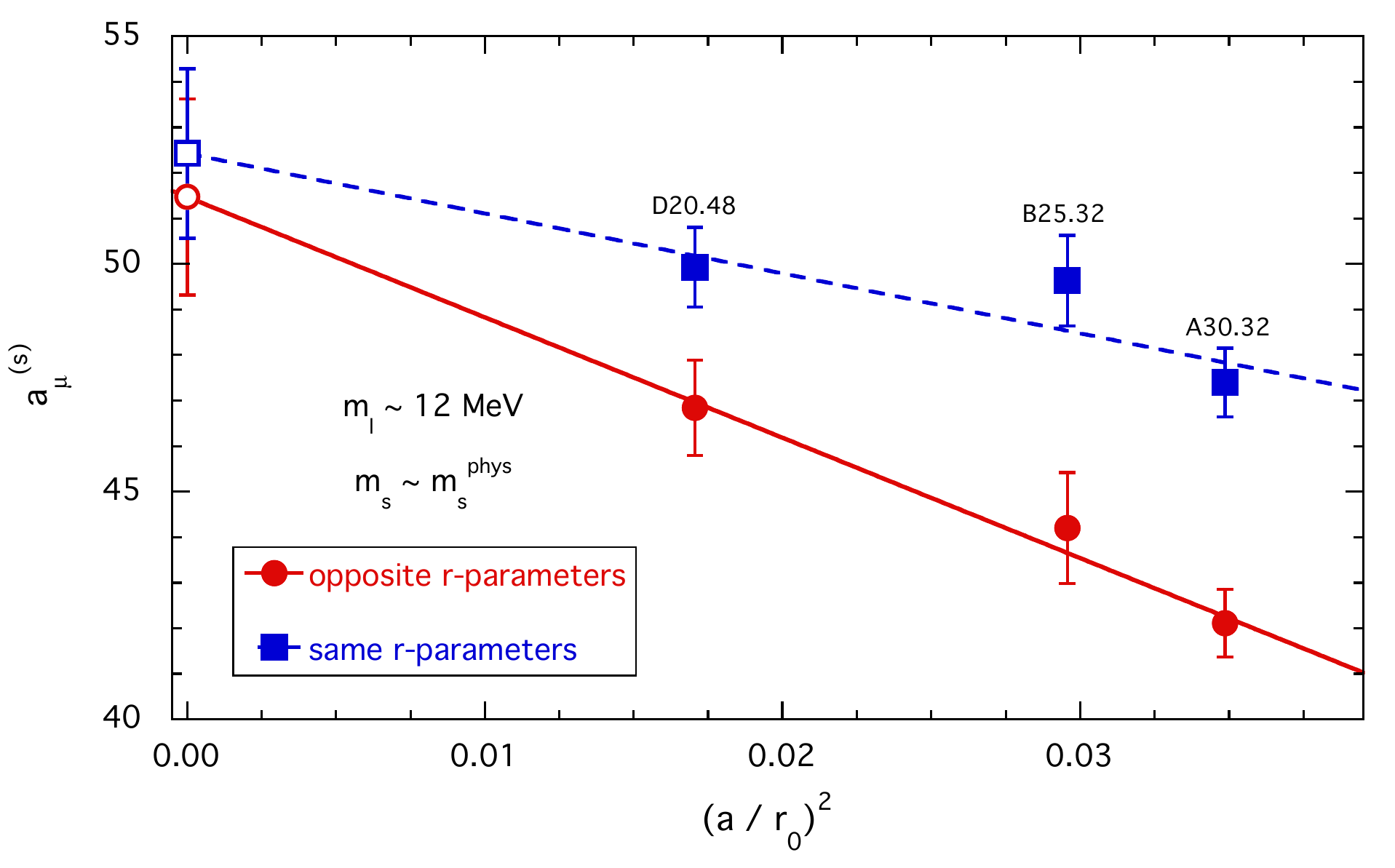}}}
\caption{\it \small Results for the strange contribution to $a_\mu^{had}$ in units of $10^{-10}$ at lowest order versus the squared lattice spacing (in units of the Sommer parameter determined in Ref.~\cite{Carrasco:2014cwa}), obtained using the local currents (\ref{eq:localVV}) (same r-parameters) and (\ref{eq:localV}) (opposite r-parameters) for the ETMC gauge ensembles A30.32, B25.32 and D20.48, which share an approximate common value of the light-quark mass and differ by the values of the lattice spacing. The empty markers represent the value extrapolated in the continuum limit assuming a linear behavior in $a^2$. }
\label{fig:amus_r}
\end{figure}

The choice (\ref{eq:localV}) differs from the one given by Eq.~(\ref{eq:localVV}) by lattice artefacts of order ${\cal{O}}(a^2)$ and by the absence of disconnected insertions.
The first point is illustrated in Fig.~\ref{fig:amus_r}, where the contribution of the strange quark to $a_\mu^{had}$, evaluated using either the current (\ref{eq:localV}) or the connected insertion of Eq.~(\ref{eq:localVV}), is shown as a function of $a^2$ for the three ensembles A30.32, B25.32 and D20.48, which share an approximate common value of the light-quark mass.
It can be seen that the same continuum limit is reached using either currents, confirming that the difference is due to discretization effects of order ${\cal{O}}(a^2)$.
Moreover, the absence of disconnected insertions in the current (\ref{eq:localV}) implies that the ``purely connected" vector correlator based on the current (\ref{eq:localVV}) is a well defined quantity and admits the hadron decomposition necessary for having the representation (\ref{eq:amu_cuts>}) (see also Refs.~\cite{Chakraborty:2015ugp,Blum:2015you} and therein quoted).

The statistical accuracy of the meson correlators is based on the use of the so-called ``one-end" stochastic method \cite{McNeile:2006bz}, which includes spatial stochastic sources at a single time slice chosen randomly.
Four stochastic sources (diagonal in the spin variable and dense in the color one) were adopted at first per each gauge configuration.

\begin{figure}[htb!]
\centering{\scalebox{0.80}{\includegraphics{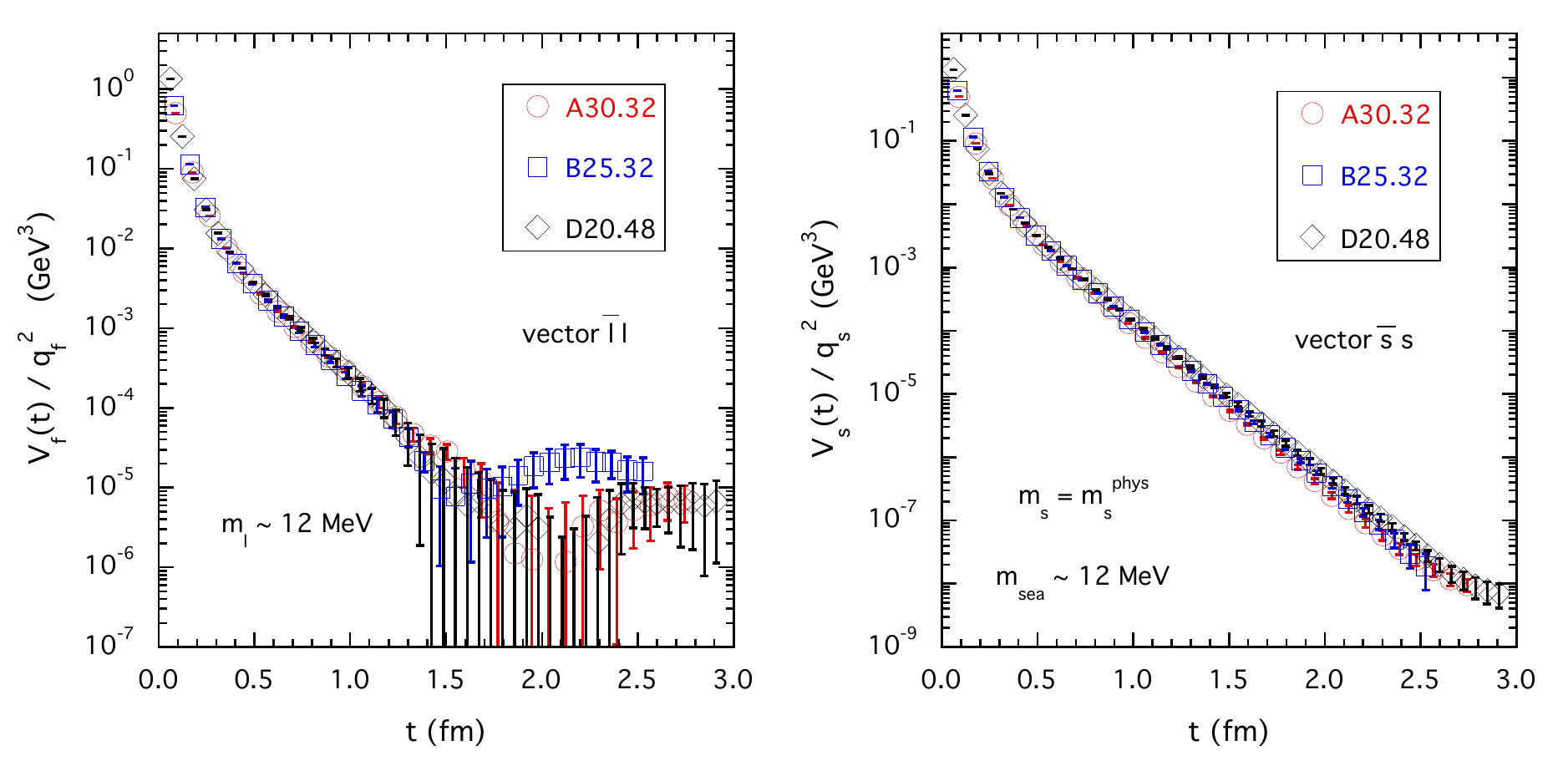}}}
\caption{\it \small The vector correlator $V(t) / q_f^2$ (in physical units) in the case of the light (left panel) and strange (right panel) quarks for the ETMC gauge ensembles specified in the inset, which share an approximate common value of the light-quark mass $m_\ell \simeq 12$ MeV and differ in the values of the lattice spacing.}
\label{fig:Vls}
\end{figure}

In the case of the light-quark contribution the signal-to-noise ratio does not allow to determine the ground-state mass $\overline{M}_V$ and the corresponding matrix element $\overline{Z}_V$ from the behavior of the vector correlator at large time distances.
This is at variance with the case of the strange and charm contributions, as it is illustrated in Fig.~\ref{fig:Vls}, where it is also shown that discretization effects are sub-leading.

\begin{figure}[htb!]
\centering{\scalebox{0.80}{\includegraphics{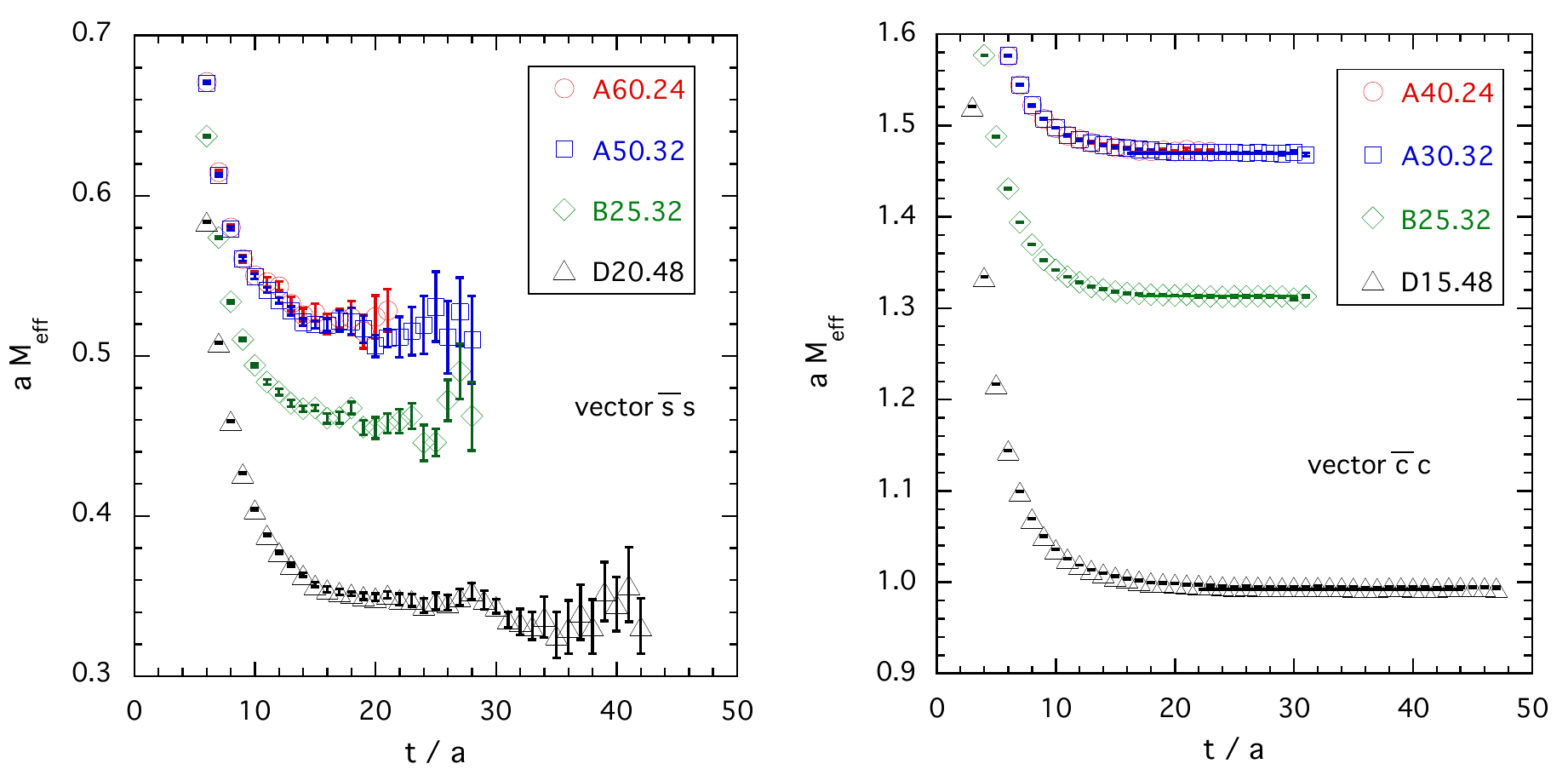}}}
\caption{\it \small Effective mass of the vector correlator $\overline{V}(\overline{t})$ in the case of the strange (left panel) and charm (right panel) contributions for the ETMC gauge ensembles specified in the insets.}
\label{fig:GS}
\end{figure}

Thus, the identification of the ground-state is presently possible only in the case of $\bar{s} s$ and $\bar{c} c$ vector mesons.
To improve the statistics we took a significative advantage by using the $\mbox{DD} - \alpha \mbox{AMG}$ solver \cite{Alexandrou:2016izb}, which has allowed us to increase by a factor of 5 the number of stochastic sources in the case of the strange quark.
In this way we find that the quality of the plateaux, shown in Fig.~\ref{fig:GS}, is acceptable in the strange sector and nice in the charm one.
In the case of the light-quark contribution an increase of the statistics by a factor $\approx 20$ is expected to be needed.

For each gauge ensemble the masses $\overline{M}_V$ and the matrix elements $\overline{Z}_V$ are extracted from a single exponential fit (including the proper backward signal) in the range $\overline{t}_{min} \leq \overline{t} \leq \overline{t}_{max}$.
The values chosen for $\overline{t}_{min}$ and $\overline{t}_{max}$ are collected in Table \ref{tab:GS}.

\begin{table}[hbt!]
\begin{center}
\begin{tabular}{||c|c||c|c||c|c||}
\hline
$\beta$ & $V / a^4$ & $\overline{t}_{min}(\bar{s} s)$ & $\overline{t}_{max}(\bar{s} s)$ & $\overline{t}_{min}(\bar{c} c)$ & 
$\overline{t}_{max}(\bar{c} c)$ \\
\hline \hline
$1.90$ & $32^3 \times 64$ &$14$ &$28$ &$16$ &$30$ \\
\cline{2-6}
            & $24^3 \times 48$ &$14$ &$20$ &$16$ &$22$ \\
\cline{2-6}
            & $20^3 \times 48$ &$14$ &$20$ &$16$ &$22$ \\
\hline \hline
$1.95$ & $32^3 \times 64$ &$15$ &$28$ &$17$ &$30$ \\
\cline{2-6}
            & $24^3 \times 48$ & $15$ &$20$ &$17$ &$22$ \\
\hline \hline
$2.10$ & $48^3 \times 96$ & $20$ &$40$ &$22$ &$44$ \\ 
\hline   
\end{tabular}
\end{center}
\vspace{-0.25cm}
\caption{\it \small Values of $\overline{t}_{min}$ and $\overline{t}_{max}$ chosen to extract the ground-state signal from the strange and charm contributions to the vector correlator $\overline{V}(\overline{t})$ for the ETMC gauge ensembles of Table \ref{tab:simudetails}.}
\label{tab:GS}
\end{table}

\section{Strange and charm contributions: lowest order}
\label{sec:s&c}

Let's start by considering the evaluation of $a_\mu^{had}(<)$ and $a_\mu^{had}(>)$ defined in Eqs.~(\ref{eq:amu_cuts<}-\ref{eq:amu_cuts>}) for various values of the ``cut'' $\overline{T}_{data}$ chosen in the range between $\overline{t}_{min}$ and  $\overline{t}_{max}$ given in Table \ref{tab:GS}.

The results for the strange contribution to $a_\mu^{had}(<)$, $a_\mu^{had}(>)$ and their sum $a_\mu^{had}$ obtained adopting four choices of $\overline{T}_{data}$, namely: $\overline{T}_{data} = (\overline{t}_{min}+2)$, $(\overline{t}_{min} + \overline{t}_{max}) / 2$, $(\overline{t}_{max} - 2)$ and $(\overline{T} / 2 - 4)$, are collected in Table \ref{tab:Ndt} for illustrative purposes in the case of few ETMC gauge ensembles.

\begin{table}[hbt!]
\begin{center}
\centerline{ensemble A40.24} 
\vspace{0.25cm}
\begin{tabular}{||c||c|c|c|c||}
\hline
$\bar{s} s$ & ~ $ (\overline{t}_{min}+2) $ ~ & ~ $ (\overline{t}_{min} + \overline{t}_{max}) / 2 $ ~ & ~ $ (\overline{t}_{max} - 2) $ ~ & ~ $ (\overline{T} /2 - 4) $ ~ \\
\hline \hline
$a_\mu^{had}(<)$ & $38.03 ~ (28)$ & $38.65 ~ (29)$ & $39.10 ~ (29)$ & $39.67 ~ (30)$ \\
\hline
$a_\mu^{had}(>)$ & $~1.97 ~ (13)$ & $~1.41 ~ (10)$ & $~1.00 ~~ (8)$ & $~0.49 ~~ (5)$ \\
\hline
$a_\mu^{had}$     & $40.00 ~ (32)$ & $40.06 ~ (31)$ & $40.10 ~ (31)$ & $40.16 ~ (31)$ \\
\hline \hline
\end{tabular}

\vspace{0.5cm}

\centerline{ensemble A30.32} 
\vspace{0.25cm}
\begin{tabular}{||c||c|c|c|c||}
\hline
$\bar{s} s$ & ~ $ (\overline{t}_{min}+2) $ ~ & ~ $ (\overline{t}_{min} + \overline{t}_{max}) / 2 $ ~ & ~ $ (\overline{t}_{max} - 2) $ ~ & ~ $ (\overline{T} /2 - 4) $ ~ \\
\hline \hline
$a_\mu^{had}(<)$ & $40.44 ~ (19)$ & $42.77 ~ (23)$ & $43.26 ~ (25)$ & $43.32 ~ (25)$ \\
\hline
$a_\mu^{had}(>)$ & $~3.15 ~ (18)$ & $~0.63 ~~ (5)$ & $~0.11 ~~ (1)$ & $~0.05 ~~ (1)$ \\
\hline
$a_\mu^{had}$     & $43.59 ~ (30)$ & $43.40 ~ (25)$ & $43.37 ~ (25)$ & $43.37 ~ (25)$ \\
\hline \hline
\end{tabular}

\vspace{0.5cm}

\centerline{ensemble B25.32} 
\vspace{0.25cm}
\begin{tabular}{||c||c|c|c|c||}
\hline
$\bar{s} s$ & ~ $ (\overline{t}_{min}+2) $ ~ & ~ $ (\overline{t}_{min} + \overline{t}_{max}) / 2 $ ~ & ~ $ (\overline{t}_{max} - 2) $ ~ & ~ $ (\overline{T} /2 - 4) $ ~ \\
\hline \hline
$a_\mu^{had}(<)$ & $40.83 ~ (14)$ & $43.18 ~ (17)$ & $44.05 ~ (18)$ & $44.16 ~ (19)$ \\
\hline
$a_\mu^{had}(>)$ & $~3.52 ~ (14)$ & $~1.11 ~~ (6)$ & $~0.23 ~~ (1)$ & $~0.11 ~~ (1)$ \\
\hline
$a_\mu^{had}$     & $44.35 ~ (22)$ & $44.29 ~ (19)$ & $44.28 ~ (19)$ & $44.27 ~ (19)$ \\
\hline \hline
\end{tabular}

\vspace{0.5cm}

\centerline{ensemble D15.48} 
\vspace{0.25cm}
\begin{tabular}{||c||c|c|c|c||}
\hline
$\bar{s} s$ & ~ $ (\overline{t}_{min}+2) $ ~ & ~ $ (\overline{t}_{min} + \overline{t}_{max}) / 2 $ ~ & ~ $ (\overline{t}_{max} - 2) $ ~ & ~ $ (\overline{T} /2 - 4) $ ~ \\
\hline \hline
$a_\mu^{had}(<)$ & $42.34 ~ (17)$ & $45.86 ~ (19)$ & $46.50 ~ (20)$ & $46.58 ~ (20)$ \\
\hline
$a_\mu^{had}(>)$ & $~4.27 ~ (18)$ & $~0.75 ~~ (5)$ & $~0.10 ~~ (1)$ & $~0.02 ~~ (1)$ \\
\hline
$a_\mu^{had}$     & $46.61 ~ (24)$ & $46.61 ~ (20)$ & $46.60 ~ (20)$ & $46.60~ (20)$ \\
\hline \hline
\end{tabular}

\end{center}
\vspace{-0.25cm}
\caption{\it \small Results for the strange contribution to $a_\mu^{had}(<)$, $a_\mu^{had}(>)$ and their sum $a_\mu^{had}$, in units of $10^{-10}$, obtained assuming $\overline{T}_{data} = (\overline{t}_{min}+2)$, $(\overline{t}_{min} + \overline{t}_{max}) / 2$, $(\overline{t}_{max} - 2)$ and $(\overline{T} / 2 - 4)$ for the ETMC gauge ensembles A40.24, A30.32, B25.32 and D15.48. Errors are statistical only.}
\label{tab:Ndt}
\end{table}

The separation between $a_\mu^{had}(<)$ and $a_\mu^{had}(>)$ depends on the specific value of $\overline{T}_{data}$, as it should be, but their sum $a_\mu^{had}$ is almost independent of the choice of the value of $\overline{T}_{data}$ in the range between $\overline{t}_{min}$ and  $\overline{t}_{max}$.
This is also reassuring of the fact that the value of $a_\mu^{had}$ is not contaminated significantly by the presence of backward signals in the correlator $\overline{V}(\overline{t})$.

In the case of the charm contribution the value of $a_\mu^{had}(>)$ is always several orders of magnitude smaller than $a_\mu^{had}(<)$ and the latter turns out to be the same for all the four choices of $\overline{T}_{data}$. 

Note that for $\overline{T}_{data} = \overline{T} / 2 - 4$ the contribution $a_\mu^{had}(>)$, which depends on the analytic representation (\ref{eq:amu_cuts>}), does not exceed $\simeq 1.2 \%$ of the total value $a_\mu^{had}$ even at the smallest value of the time extension $\overline{T}$.

In what follows all the four choices of $\overline{T}_{data}$ will be employed in the various branches of our bootstrap analysis.
The corresponding systematics is largely sub-dominant with respect to the other sources of uncertainties and it will not be given separately in the error budget.

\begin{figure}[htb!]
\centering{\scalebox{0.80}{\includegraphics{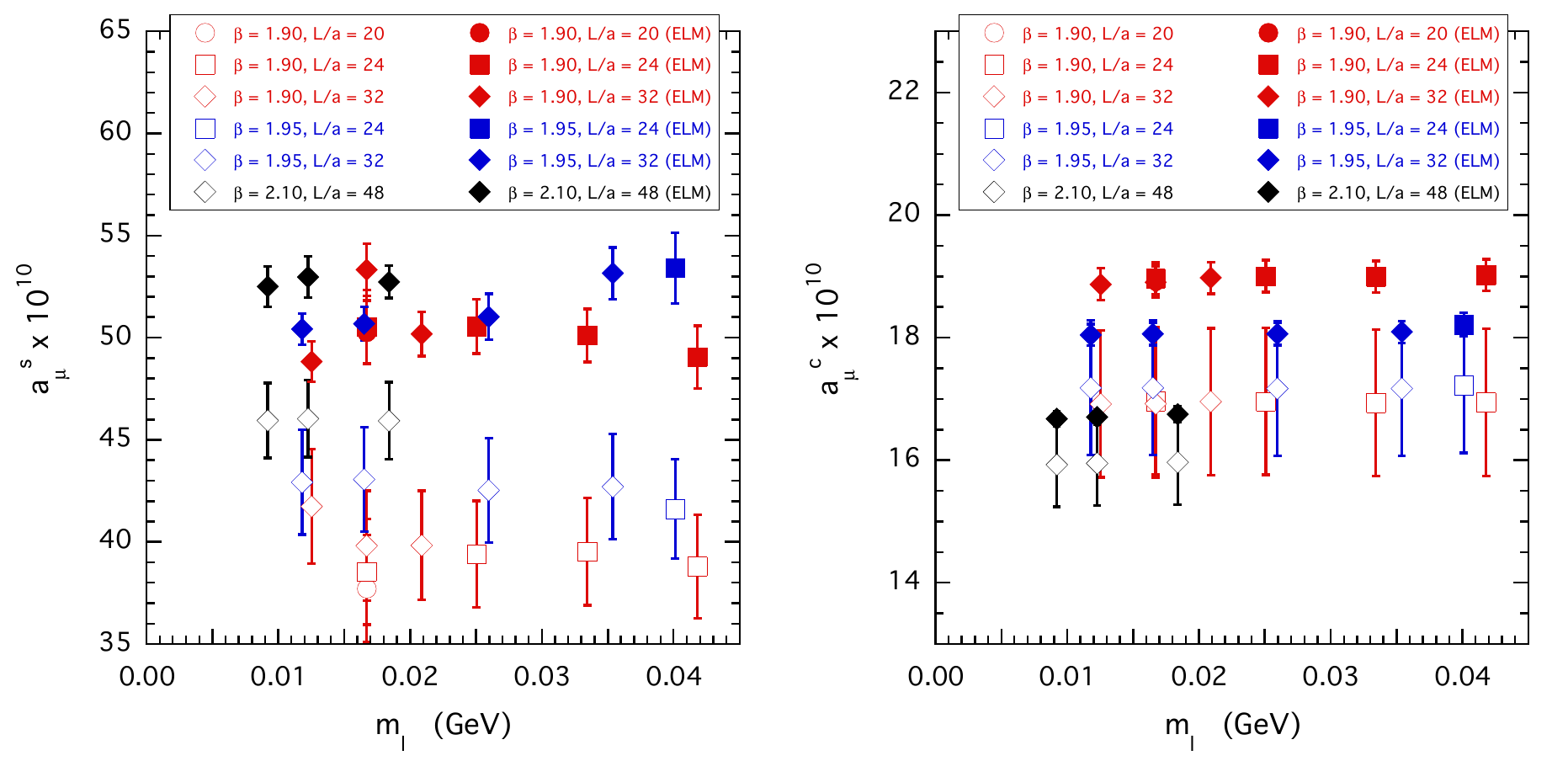}}}
\caption{\it \small Results for the strange (left panel) and charm (right panel) contributions to $a_\mu^{had}$ in units of $10^{-10}$, evaluated with (filled markers) and without (empty markers) the ELM procedure described by Eq.~(\ref{eq:HLtrick}). The PDG values $M_V^{(phys)} = 1.0195$ and $3.0969$ GeV \cite{PDG} have been adopted for the physical $\bar{s} s$ and $\bar{c} c$ vector meson masses, respectively.}
\label{fig:amuq}
\end{figure}

The results obtained for the strange and charm contributions to $a_\mu^{had}$ are shown by the empty markers in Fig.~\ref{fig:amuq}.
We observe a mild dependence on the light-quark mass, being driven only by sea quarks, and also small residual FSEs visible only in the case of the strange contribution.
The errors of the data turn out to be dominated by the uncertainties of the scale setting, which are similar for all the gauge ensembles used in this work.

In Ref.~\cite{Burger:2013jya} a modification of the calculated $a_\mu^{had}$ at pion masses above the physical point has been proposed in order to weaken the pion mass dependence of the resulting $a_\mu^{had}$ for improving the reliability of the chiral extrapolation. 
Though the procedure of Ref.~\cite{Burger:2013jya} has been conceived mainly for the light contribution to $a_\mu^{had}$, we have explored its usefulness also in the case of the strange and charm contributions.
The proposal consists in multiplying the Euclidean 4-momentum transfer $Q^2$ by a factor equal to $(M_V / M_V^{phys})^2$ in order to modify the $Q^2$-dependence of the HVP function $\Pi_R(Q^2)$ without modifying its value at the physical point.
One obtains the same effect in our master formulae by redefining the lepton mass as
 \be
    \overline{m}_\mu^{ELM} = \overline{M}_V \frac{m_\mu}{M_V^{phys}} ~ .
    \label{eq:HLtrick}
 \ee
The expected advantage of the use of the effective lepton mass (\ref{eq:HLtrick}) comes from the fact that the kernel function, and therefore $a_\mu^{had}$, depends only on the lepton mass in lattice units (see Eq.~(\ref{eq:ftau})).
Thanks to Eq.~(\ref{eq:HLtrick}), which will be referred to as the Effective Lepton Mass (ELM) procedure, the knowledge of the value of the lattice spacing is not required and therefore the resulting $a_\mu^{had}$ is not affected by the uncertainties of the scale setting.
The drawback of the ELM procedure is instead represented by its potential sensitivity to the statistical fluctuations of the vector meson mass extracted from the lattice data.

The results obtained adopting the ELM procedure (\ref{eq:HLtrick}) in the case of the strange and charm contributions to $a_\mu^{had}$ are shown by the filled markers in Fig.~\ref{fig:amuq}, where the physical values for the $\bar{s} s$ and $\bar{c} c$ vector masses have been taken from PDG \cite{PDG} (namely, $M_V^{(phys)} = 1.0195$ and $3.0969$ GeV, respectively)\footnote{We have checked that the chiral and continuum extrapolations of the simulated vector meson masses are consistent with the PDG values within lattice uncertainties, which are dominated by the error of the lattice scale.}.
It can be seen that the ELM procedure reduces remarkably the overall uncertainty of the data.
Moreover, it further weakens the pion mass dependence (in any case driven only by the sea quarks) and modifies the discretization effects, leading to a better scaling behavior of the data in the case of the charm contribution.
Since the pion mass dependence is in any case quite mild, the ELM procedure can be viewed as an alternative way to perform the continuum extrapolation and to avoid the scale setting uncertainties.

\begin{figure}[htb!]
\centering{\scalebox{0.80}{\includegraphics{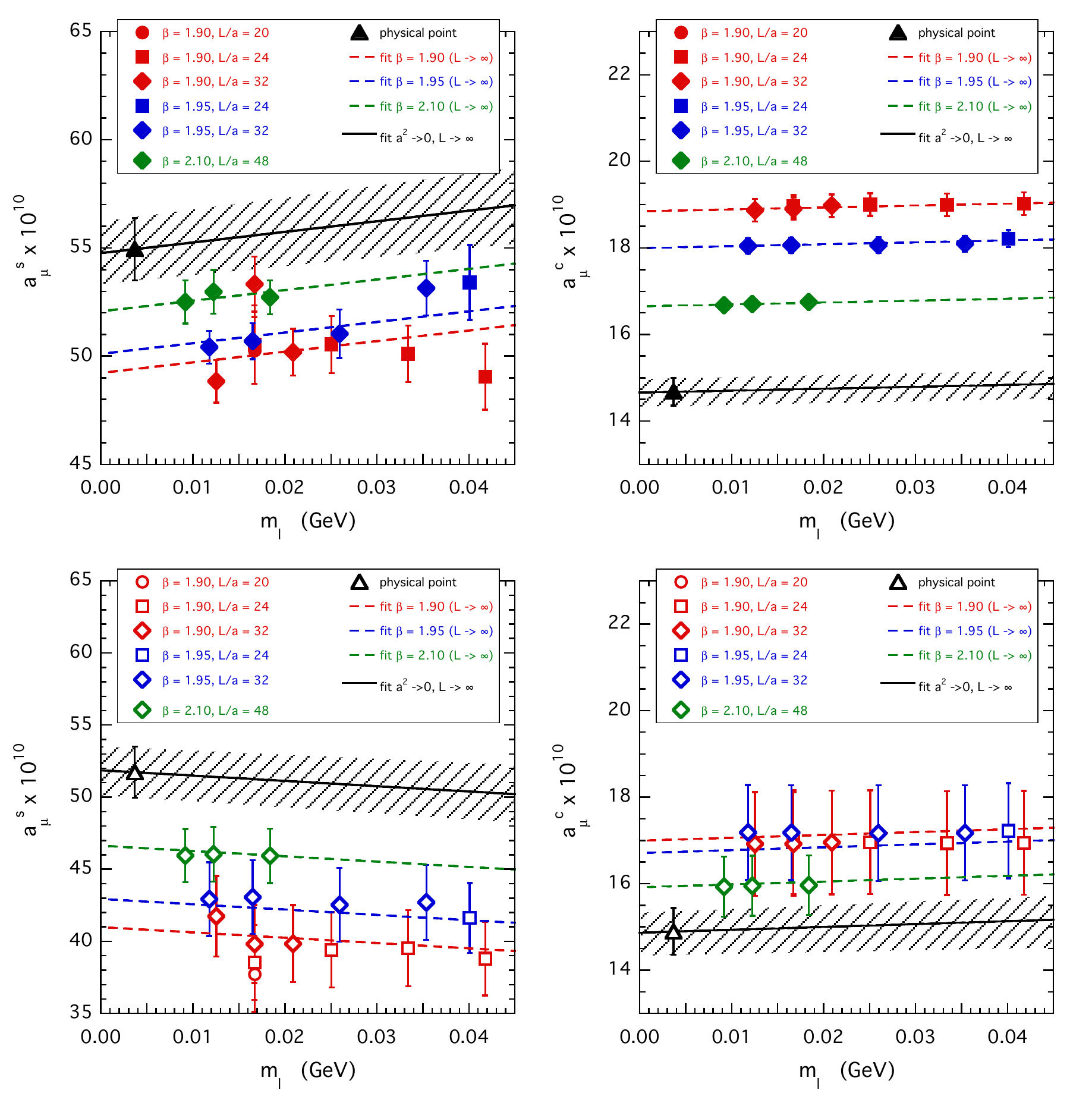}}}
\caption{\it \small Results for the strange (left panels) and charm (right panels) contributions to $a_\mu^{had}$ in units of $10^{-10}$ for the ETMC gauge ensembles of Table \ref{tab:simudetails}. Upper (lower) panels correspond to the data obtained with (without) the ELM procedure (\ref{eq:HLtrick}). The dashed lines correspond to the linear fit (\ref{eq:fit_amus}) including the discretization term in the infinite volume limit. The solid lines correspond to the linear fit (\ref{eq:fit_amus}) in the continuum and infinite volume limits, while the shaded areas identify the corresponding uncertainty at the level of one standard deviation. The triangles are the results of the extrapolation at the physical pion mass and in the continuum and infinite volume limits based on the linear fit (\ref{eq:fit_amus}).}
\label{fig:amuq_fit}
\end{figure}

Using the data obtained either with or without the ELM procedure we have performed a combined fit for the extrapolation to the physical pion mass, the continuum and infinite volume limits using the following simple Ansatz
 \be
     a_\mu^{s,c} = A_0^{s,c} \left[ 1 + A_1^{s,c} \xi + D^{s,c} a^2 + F^{s,c} \xi \frac{e^{-M_\pi L}}{M_\pi L} \right] ~ ,
     \label{eq:fit_amus}
  \ee
where $\xi \equiv M_\pi^2 / (4 \pi f_0)^2$ and the exponential term is a phenomenological representation of possible finite size effects (FSEs).
The results of the linear fit (\ref{eq:fit_amus}) are shown in Fig.~\ref{fig:amuq_fit} by the solid lines.
In our combined fit the values of the parameters are determined by a $\chi^2$-minimization procedure adopting an uncorrelated $\chi^2$. 
The uncertainties on the fitting parameters do not depend on the value of the uncorrelated $\chi^2$, because they are obtained using the bootstrap procedure of Ref.~\cite{Carrasco:2014cwa} (see Section~\ref{sec:simulations}).
This guarantees that all correlations among the lattice data points and among the fitting parameters are properly taken into account.

Averaging over the results corresponding to different fitting functions of the data either with or without the ELM procedure we get at the physical point
 \bea
     a_\mu^{s, phys} & = & (53.1 \pm 1.6_{stat+fit} \pm 1.5_{input} \pm 1.3_{disc} \pm 0.2_{FSE} \pm 0.1_{chir}) \cdot 10^{-10} ~ , \nonumber \\
                               & = & (53.1 \pm 1.6_{stat+fit} \pm 2.0_{syst}) \cdot 10^{-10} ~ , \nonumber \\
                               & = & (53.1 \pm 2.5) \cdot 10^{-10} ~ ,
     \label{eq:amus_phys}
 \eea
where
\begin{itemize}
\item $()_{stat+fit}$ indicates the uncertainty induced by both the statistical errors and the fitting procedure itself;
\item $()_{input}$ is the error coming from the uncertainties of the input parameters of the eight branches of the quark mass analysis of Ref.~\cite{Carrasco:2014cwa};
\item $()_{disc}$ is the uncertainty due to both discretization effects and scale setting, estimated by comparing the results obtained with and without the ELM procedure (\ref{eq:HLtrick});
\item $()_{FSE}$ is the error coming from including ($F^s \neq 0$) or excluding ($F^s = 0$) the FSE correction. When FSEs are not included, all the gauge ensembles with $L / a = 20$ and $24$ are also not included;
\item $()_{chir}$ is the error coming from including ($A_1^s \neq 0$) or excluding ($A_1^s = 0$) the linear term in the light-quark mass.
\end{itemize}

Our result (\ref{eq:amus_phys}) compares well with the $N_f = 2+1+1$  result $a_\mu^{s, phys} = (53.41 \pm 0.59) \cdot 10^{-10}$ from the HPQCD collaboration~\cite{Chakraborty:2014mwa}, the $N_f = 2+1$ finding $a_\mu^{s, phys} = (53.1 \pm 0.9_{-0.3}^{+0.1}) \cdot 10^{-10}$ obtained by the RBC/UKQCD collaboration~\cite{Blum:2016xpd}, and with the recent $N_f = 2$ result $a_\mu^{s, phys} = (51.1 \pm 1.7 \pm 0.4) \cdot 10^{-10}$ of Ref.~\cite{DellaMorte:2017dyu}.

In the case of the charm contribution we obtain
 \bea
     a_\mu^{c, phys} & = & (14.75 \pm 0.42_{stat+fit} \pm 0.36_{input} \pm 0.10_{disc} \pm 0.03_{FSE} \pm 0.01_{chir}) \cdot 10^{-10} ~ , \nonumber \\
                                & = & (14.75 \pm 0.42_{stat+fit} \pm 0.37_{syst}) \cdot 10^{-10} ~ , \nonumber \\
                                & = & (14.75 \pm 0.56) \cdot 10^{-10} ~ ,
     \label{eq:amuc_phys}
 \eea
where the errors are estimated as in the case of the strange quark contribution.
Our finding (\ref{eq:amuc_phys}) agrees with the $N_f = 2+1+1$ result $a_\mu^{c, phys} = (14.42 \pm 0.39) \cdot 10^{-10}$ from the HPQCD collaboration~\cite{Chakraborty:2014mwa} and with recent $N_f = 2$ one $a_\mu^{c, phys} = (14.3 \pm 0.2 \pm 0.1) \cdot 10^{-10}$ of Ref.~\cite{DellaMorte:2017dyu}.

\section{Strange and charm contributions: e.m.~corrections}
\label{sec:deltas&c}

\begin{figure}[htb!]
\begin{center}
\includegraphics[scale=1.8]{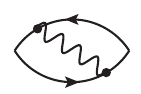} ~ 
\includegraphics[scale=1.8]{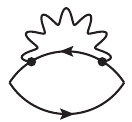} ~ 
\includegraphics[scale=1.8]{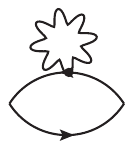} ~ 
\includegraphics[scale=1.8]{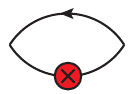} ~ 
\includegraphics[scale=1.8]{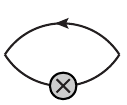}\\
~ (a) \qquad \qquad \qquad (b) \qquad \qquad \qquad ~ (c) \qquad \qquad \qquad ~ (d) \qquad \qquad \qquad (e) ~
\end{center}
\caption{\it Fermionic connected diagrams contributing to the e.m.~corrections to $a_\mu^{had}$: exchange (a), self energy (b), tadpole (c), pseudoscalar (d) and scalar (e) insertions. Solid lines represent quark propagators.}
\label{fig:diagrams}
\end{figure}

Let's now turn to the e.m.~corrections at leading order in $\alpha_{em}$ to $a_\mu^{had}$, which using the expansion method of Ref.~\cite{deDivitiis:2013xla} require the evaluation of the self-energy, exchange, tadpole, pseudoscalar and scalar insertion diagrams depicted in Fig.~\ref{fig:diagrams}.

For each quark flavor $f$ the e.m.~correction $\delta V(t)$ to the vector correlator is given by
 \be
     \delta V(t) \equiv \delta V^{self}(t) + \delta V^{exch}(t) + \delta V^{tad}(t) + \delta V^{PS}(t) + \delta V^S(t) ~ 
     \label{eq:deltaV}
 \ee
with
 \bea
     \label{eq:deltaV_SE}
     \delta V^{self+exch}(t) & = & \frac{4 \pi \alpha_{em}}{3} \sum_{i = 1, 2, 3} ~ \sum_{\vec{x}, y_1, y_2} \langle 0| T \left \{ J_i^\dagger(\vec{x}, t) ~ 
                                                   \sum_\mu J_\mu^C(y_1) J_\mu^C(y_2) ~ J_i(0) \right \} | 0 \rangle ~ , \qquad \\
    \label{eq:deltaV_T}
     \delta V^{tad}(t) & = &  \frac{4 \pi \alpha_{em}}{3} \sum_{i = 1, 2, 3} ~ \sum_{\vec{x}, y} ~ \langle 0| T \left \{ J_i^\dagger(\vec{x}, t) ~ 
                                          \sum_\nu T_\nu(y) ~ J_i(0) \right \} | 0 \rangle ~ , \\
     \label{eq:deltaV_PS}
     \delta V^{PS}(t) & = & \frac{2 \delta m_f^{crit}}{3} \sum_{i = 1, 2, 3} ~ \sum_{\vec{x}, y} ~ \langle 0| T \left \{ J_i^\dagger(\vec{x}, t) ~ 
                                         i \overline{\psi}_f(y) \gamma_5  \psi_f(y) ~ J_i(0) \right \} | 0 \rangle ~ , \\
     \label{eq:deltaV_S}
     \delta V^S(t) & = & - \frac{2 m_f}{3 Z_m {\cal{Z}}_f} \sum_{i = 1, 2, 3} ~ \sum_{\vec{x}, y} ~ \langle 0| T \left \{ J_i^\dagger(\vec{x}, t) ~ 
                                    \overline{\psi}_f(y)  \psi_f(y) ~ J_i(0) \right \} | 0 \rangle ~ , 
 \eea
where $J_\mu^C(y)$ and $T_\nu(y)$ are given in Eqs.~(\ref{eq:conservedV}) and (\ref{eq:tadpole}), respectively. 
In Eq.~(\ref{eq:deltaV}) $\delta m_f^{crit}$ is the e.m.~shift of the critical mass for the quark flavor $f$, while $Z_m$ and ${\cal{Z}}_f$ are related to the mass renormalization constants (RCs) in QCD and QCD+QED.
For our maximally twisted-mass setup $\delta m_f^{crit}$ has been determined in Ref.~\cite{Giusti:2017dmp}, while $1 / Z_m = Z_P$, where $Z_P$ is the RC of the pseudoscalar density evaluated in Ref.~\cite{Carrasco:2014cwa}.
For $1 / {\cal{Z}}_f$ we use the perturbative result at leading order in $\alpha_{em}$ in the $\overline{MS}$ scheme, given by \cite{Martinelli:1982mw,Aoki:1998ar}
 \be
      \frac{1}{{\cal{Z}}_f}(\overline{\mbox{MS}}, \mu) = \frac{\alpha_{em} q_f^2}{4 \pi} \left[ 6 \mbox{log}(a \mu) - 22.5954 \right] ~ ,
      \label{eq:Zf}
 \ee
where the renormalization scale $\mu$ is taken to be equal to $\mu = 2 $ GeV, at which we consider that the renormalized quark masses in QCD and QCD+QED coincide (see Ref.~\cite{Giusti:2017dmp}). 

The removal of the photon zero-mode is done according to $QED_L$~\cite{Hayakawa:2008an}, i.e.~the photon field $A_\mu$ satisfies $A_\mu(k_0, \vec{k} = \vec{0}) \equiv 0$ for all $k_0$.

Within the quenched QED approximation, which neglects the effects of the sea-quark electric charges, the correlator $\delta V^{self}(t) + \delta V^{exch}(t)$ corresponds to the sum of the diagrams (\ref{fig:diagrams}a)-(\ref{fig:diagrams}b), while the correlators $\delta V^{tad}(t)$, $\delta V^{PS}(t)$ and $\delta V^S(t)$ represent the contributions of the diagrams (\ref{fig:diagrams}c), (\ref{fig:diagrams}d) and (\ref{fig:diagrams}e), respectively.
In the quenched QED approximation the shift $\delta m_f^{crit}$ is proportional to $\alpha_{em} q_f^2$ (see for details Ref.~\cite{Giusti:2017dmp}).
 
In addition one has to consider also the QED contribution to the renormalization constant of the vector current (\ref{eq:localV}), namely
 \be
     Z_A = Z_A^{(0)} + \delta Z_A + {\cal{O}}(\alpha_{em}^2) ~ ,
     \label{eq:ZAem}
 \ee
where $Z_A^{(0)}$ is the renormalization constant (RC) of the current in absence of QED (determined in Ref.~\cite{Carrasco:2014cwa}) and $\delta Z_A$ is the ${\cal{O}}(\alpha_{em})$ RC.
The latter can be written as
 \be
     \delta Z_A = Z_A^{(0)} \cdot Z_A^{(em)} \cdot Z_A^{(fact)} ~ ,
     \label{eq:deltaZA}
 \ee
where $Z_A^{(em)}$ is the one-loop perturbative estimate of the QED effect at order ${\cal{O}}(\alpha_s^0)$ in the strong coupling and $Z_A^{(fact)}$ takes into account corrections of order ${\cal{O}}(\alpha_{em} \alpha_s^n)$ with $n \geq 1$, i.e.~corrections to the ``naive factorization'' approximation in which $Z_A^{(fact)} = 1$.
In the Appendix \ref{sec:ZA} we present our non-perturbative estimate $Z_A^{(fact)} = 0.9 \pm 0.1$, obtained through the use of the axial Ward-Takahashi identity (WTI) derived in the presence of QED effects\footnote{A different non-perturbative procedure for evaluating the QED contribution to the RC of the local vector current has been recently developed in Ref.~\cite{Boyle:2017gzv}.}.
Using the result $Z_A^{(em)} = - 15.7963 ~ \alpha_{em} ~ q_f^2 / (4 \pi)$ from Refs.~\cite{Martinelli:1982mw,Aoki:1998ar}, we have to add to Eq.~(\ref{eq:deltaV}) the following contribution
 \be
      \delta V^{Z_A}(t) \equiv -2.51406 ~ \alpha_{em} q_f^2 ~ Z_A^{(fact)} ~ V(t) ~ .
      \label{eq:deltaV_ZA}
 \ee

Thus, the e.m.~corrections $\delta a_\mu^{had}$ can be written as
\be
     \delta a_\mu^{had} \equiv \delta a_\mu^{had}(<) + \delta a_\mu^{had}(>)
     \label{eq:deltamu}
 \ee
with (see Eqs.~(\ref{eq:amu_cuts<}-\ref{eq:amu_cuts>}))
 \bea 
     \label{eq:deltamu_cuts<}
     \delta a_\mu^{had}(<) & = & 4 \alpha_{em}^2 \sum_{\overline{t} = 0}^{\overline{T}_{data}} \overline{f}(\overline{t}) ~ \delta \overline{V}(\overline{t}) ~ , \\
     \label{eq:deltamu_cuts>}
     \delta a_\mu^{had}(>) & = & 4 \alpha_{em}^2 \sum_{\overline{t} = \overline{T}_{data} + 1}^\infty \overline{f}(\overline{t}) ~ 
                                                  \delta \left[ \frac{\overline{Z}_V} {2 \overline{M}_V} e^{- \overline{M}_V \overline{t}} \right] \nonumber \\
                                         & = & 4 \alpha_{em}^2 \sum_{\overline{t} = \overline{T}_{data} + 1}^\infty \overline{f}(\overline{t}) ~ 
                                                   \frac{\overline{Z}_V} {2 \overline{M}_V} e^{- \overline{M}_V \overline{t}} \left[ \frac{\delta \overline{Z}_V}{\overline{Z}_V}
                                                  - \frac{\delta \overline{M}_V}{\overline{M}_V} (1 + \overline{M}_V \overline{t}) \right] , \qquad 
 \eea
where $\delta \overline{M}_V$ and $\delta \overline{Z}_V$ can be determined, respectively, from the ``slope'' and the ``intercept'' of the ratio $\delta \overline{V}(\overline{t}) / \overline{V}(\overline{t})$ at large time distances $\overline{t}_{min} \leq \overline{t} \leq \overline{t}_{max}$ (see Refs.~\cite{deDivitiis:2011eh,deDivitiis:2013xla,Giusti:2017dmp}).
Note that all the quantities $\delta \overline{V}$, $\delta \overline{Z}_V$ and $\delta \overline{M}_V$ are proportional to $\alpha_{em} q_f^2$, which make $\delta a_\mu^{had}$ proportional to $\alpha_{em}^3 q_f^4$.

\begin{figure}[htb!]
\centering{\scalebox{0.80}{\includegraphics{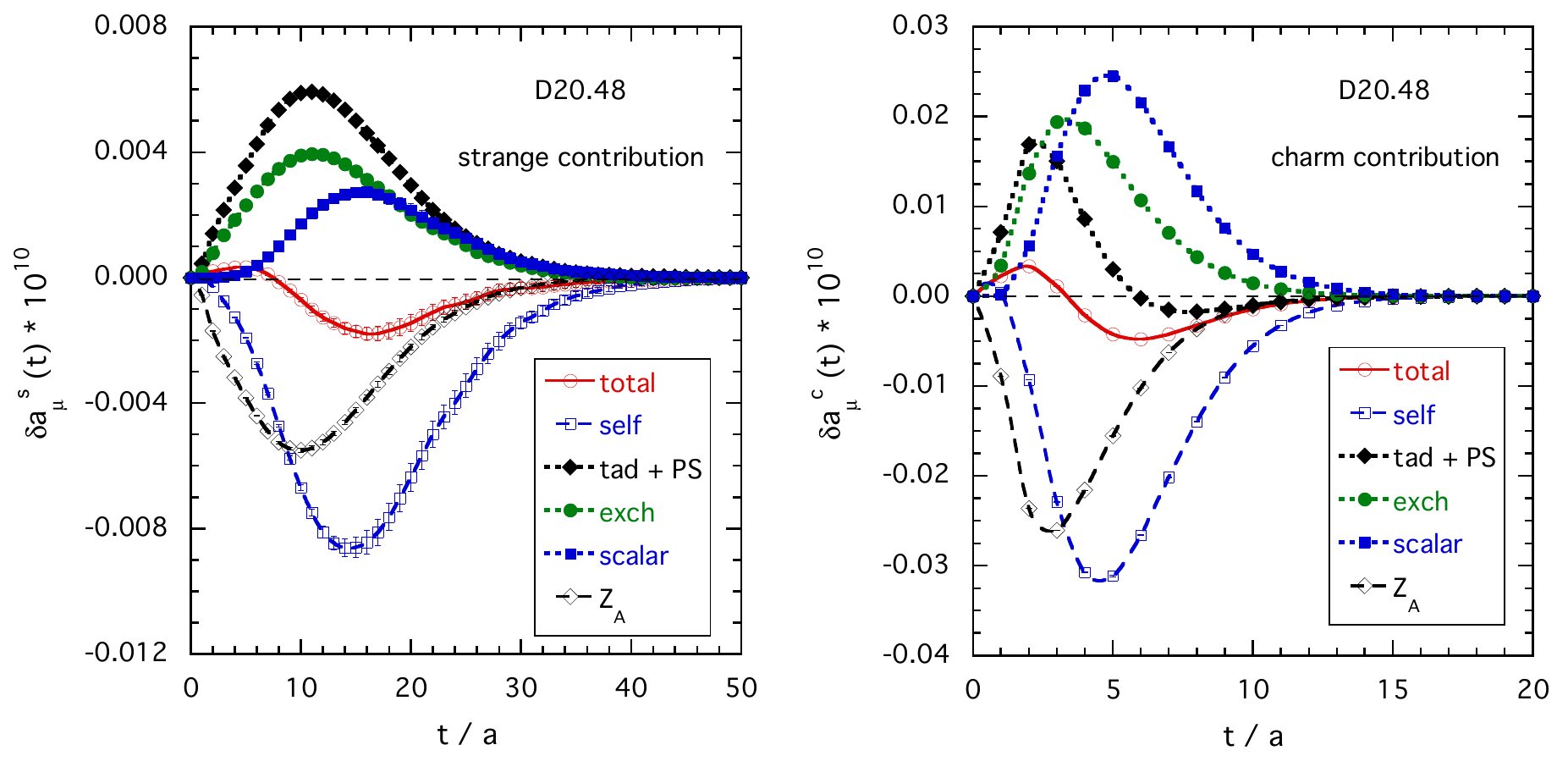}}}
\caption{\it \small The time behavior of the integrand function $\delta a_\mu^{had}(t)$ in the r.h.s.~of Eqs.~(\ref{eq:deltamu_cuts<}-\ref{eq:deltamu_cuts>}) in the case of the strange (left panel) and charm (right panel) quarks in units of $10^{-10}$, obtained for the ETMC gauge ensemble D20.48. The labels ``self'', ``tad + PS'', ``exch'', ``scalar'' and ``$Z_A$'' indicate the contributions of the diagrams (\ref{fig:diagrams}b),  (\ref{fig:diagrams}c)+(\ref{fig:diagrams}d),  (\ref{fig:diagrams}a), (\ref{fig:diagrams}e) and the one generated by the QED effect in the RC $Z_A$ of the local vector current at leading order in $\alpha_{em}$ (see Eq.~(\ref{eq:deltaV_ZA})) with $Z_A^{(fact)} = 0.9$. The label ``total'' corresponds to the sum of all the contributions.}
\label{fig:damuqt}
\end{figure}

The time dependence of the integrand function in the r.h.s.~of Eqs.~(\ref{eq:deltamu_cuts<}-\ref{eq:deltamu_cuts>}) is shown in Fig.~\ref{fig:damuqt} in the case of the ETMC gauge ensemble D20.48.
The contributions coming from the various diagrams of Fig.~\ref{fig:diagrams} as well as from the additional term (\ref{eq:deltaV_ZA}) are determined quite precisely and are characterized by different signs.
Partial cancellations among the various contributions occur in the total sum, which turns out to be smaller than each individual contributions. 
Thus, even a $10 \%$ uncertainty on the RC $\delta Z_A$ may have a larger impact on the final uncertainty of $ \delta a_\mu^{had}$, as it will be shown later on.

The results for the strange contribution to $\delta a_\mu^{had}(<)$, $\delta a_\mu^{had}(>)$ and their sum $\delta a_\mu^{had}$, obtained adopting the four choices of $\overline{T}_{data}$, namely: $\overline{T}_{data} = (\overline{t}_{min}+2)$, $(\overline{t}_{min} + \overline{t}_{max}) / 2$, $(\overline{t}_{max} - 2)$ and $(\overline{T} / 2 - 4)$, are collected in Table \ref{tab:deltaNdt} for some of the ETMC gauge ensembles.

\begin{table}[hbt!]
\begin{center}

\centerline{ensemble A40.24} 
\vspace{0.25cm}
\begin{tabular}{||c||c|c|c|c||}
\hline
$\bar{s} s$ & ~ $ (\overline{t}_{min}+2) $ ~ & ~ $ (\overline{t}_{min} + \overline{t}_{max}) / 2 $ ~ & ~ $ (\overline{t}_{max} - 2) $ ~ & ~ $ (\overline{T} /2 - 4) $ ~ \\
\hline \hline
$\delta a_\mu^{had}(<)$ & $-1.26~ (13)$ & $-1.36 ~ (14)$ & $-1.45 ~ (14)$ & $-1.58 ~ (15)$ \\
\hline
$\delta a_\mu^{had}(>)$ & $-0.40 ~~ (7)$ & $-0.31 ~~ (6)$ & $-0.24 ~~ (5)$ & $-0.13 ~~ (3)$ \\
\hline
$\delta a_\mu^{had}$     & $-1.66 ~ (16)$ & $-1.67 ~ (16)$ & $-1.69 ~ (16)$ & $-1.71 ~ (16)$ \\
\hline \hline
\end{tabular}

\vspace{0.5cm}

\centerline{ensemble A30.32} 
\vspace{0.25cm}
\begin{tabular}{||c||c|c|c|c||}
\hline
$\bar{s} s$ & ~ $ (\overline{t}_{min}+2) $ ~ & ~ $ (\overline{t}_{min} + \overline{t}_{max}) / 2 $ ~ & ~ $ (\overline{t}_{max} - 2) $ ~ & ~ $ (\overline{T} /2 - 4) $ ~ \\
\hline \hline
$\delta a_\mu^{had}(<)$ & $-1.03 ~ (10)$ & $-1.44 ~ (15)$ & $-1.56 ~ (17)$ & $-1.58 ~ (17)$ \\
\hline
$\delta a_\mu^{had}(>)$ & $-0.56 ~~ (8)$ & $-0.16 ~~ (3)$ & $-0.03 ~~ (1)$ & $-0.02 ~~ (1)$ \\
\hline
$\delta a_\mu^{had}$     & $-1.59 ~ (18)$ & $-1.60 ~ (17)$ & $-1.59 ~ (17)$ & $-1.60 ~ (18)$ \\
\hline \hline
\end{tabular}

\vspace{0.5cm}

\centerline{ensemble B25.32} 
\vspace{0.25cm}
\begin{tabular}{||c||c|c|c|c||}
\hline
$\bar{s} s$ & ~ $ (\overline{t}_{min}+2) $ ~ & ~ $ (\overline{t}_{min} + \overline{t}_{max}) / 2 $ ~ & ~ $ (\overline{t}_{max} - 2) $ ~ & ~ $ (\overline{T} /2 - 4) $ ~ \\
\hline \hline
$\delta a_\mu^{had}(<)$ & $-1.35 ~ (12)$ & $-1.80 ~ (15)$ & $-2.05 ~ (17)$ & $-2.09 ~ (18)$ \\
\hline
$\delta a_\mu^{had}(>)$ & $-0.80 ~~ (8)$ & $-0.34 ~~ (4)$ & $-0.09 ~~ (1)$ & $-0.05 ~~ (1)$ \\
\hline
$\delta a_\mu^{had}$     & $-2.15 ~ (18)$ & $-2.14 ~ (18)$ & $-2.14 ~ (18)$ & $-2.14 ~ (18)$ \\
\hline \hline
\end{tabular}

\vspace{0.5cm}

\centerline{ensemble D15.48} 
\vspace{0.25cm}
\begin{tabular}{||c||c|c|c|c||}
\hline
$\bar{s} s$ & ~ $ (\overline{t}_{min}+2) $ ~ & ~ $ (\overline{t}_{min} + \overline{t}_{max}) / 2 $ ~ & ~ $ (\overline{t}_{max} - 2) $ ~ & ~ $ (\overline{T} /2 - 4) $ ~ \\
\hline \hline
$\delta a_\mu^{had}(<)$ & $-1.27 ~~ (9)$ & $-1.86 ~ (15)$ & $-2.02 ~ (18)$ & $-2.04 ~ (19)$ \\
\hline
$\delta a_\mu^{had}(>)$ & $-0.77 ~ (13)$ & $-0.19 ~~ (4)$ & $-0.03 ~~ (1)$ & $-0.01 ~~ (1)$ \\
\hline
$\delta a_\mu^{had}$     & $-2.04 ~ (20)$ & $-2.05 ~ (19)$ & $-2.05 ~ (19)$ & $-2.05 ~ (19)$ \\
\hline \hline
\end{tabular}

\end{center}
\vspace{-0.25cm}
\caption{\it \small Results for the strange contribution to $\delta a_\mu^{had}(<)$, $\delta a_\mu^{had}(>)$ and their sum $\delta a_\mu^{had}$, in units of $10^{-12}$, obtained assuming $\overline{T}_{data} = (\overline{t}_{min}+2)$, $(\overline{t}_{min} + \overline{t}_{max}) / 2$, $(\overline{t}_{max} - 2)$ and $(\overline{T} / 2 - 4)$ for the ETMC gauge ensembles A40.24, A30.32, B25.32 and D15.48. Errors are statistical only.}
\label{tab:deltaNdt}
\end{table}

As in the case of the lowest-order terms $a_\mu^{had}(<)$ and $a_\mu^{had}(>)$, we find that the separation between $\delta a_\mu^{had}(<)$ and $\delta a_\mu^{had}(>)$ depends on the specific value of $\overline{T}_{data}$, as it should be, but their sum $\delta a_\mu^{had}$ is largely independent of the choice of the value of $\overline{T}_{data}$ in the range between $\overline{t}_{min}$ and  $\overline{t}_{max}$ within the statistical uncertainties.
As in the case of the lowest-order term, the contribution $\delta a_\mu^{had}(>)$, which depends on the analytic representation (\ref{eq:deltamu_cuts>}), is significantly reduced at $\overline{T}_{data} = \overline{T} / 2 - 4$, where it does not exceed the statistical uncertainty of $\delta a_\mu^{had}$ .

In the case of the charm contribution the value of $\delta a_\mu^{had}(>)$ is always several orders of magnitude smaller than $\delta a_\mu^{had}(<)$ and the latter turns out to be the same for all the four choices of $\overline{T}_{data}$. 

\begin{figure}[htb!]
\centering{\scalebox{0.80}{\includegraphics{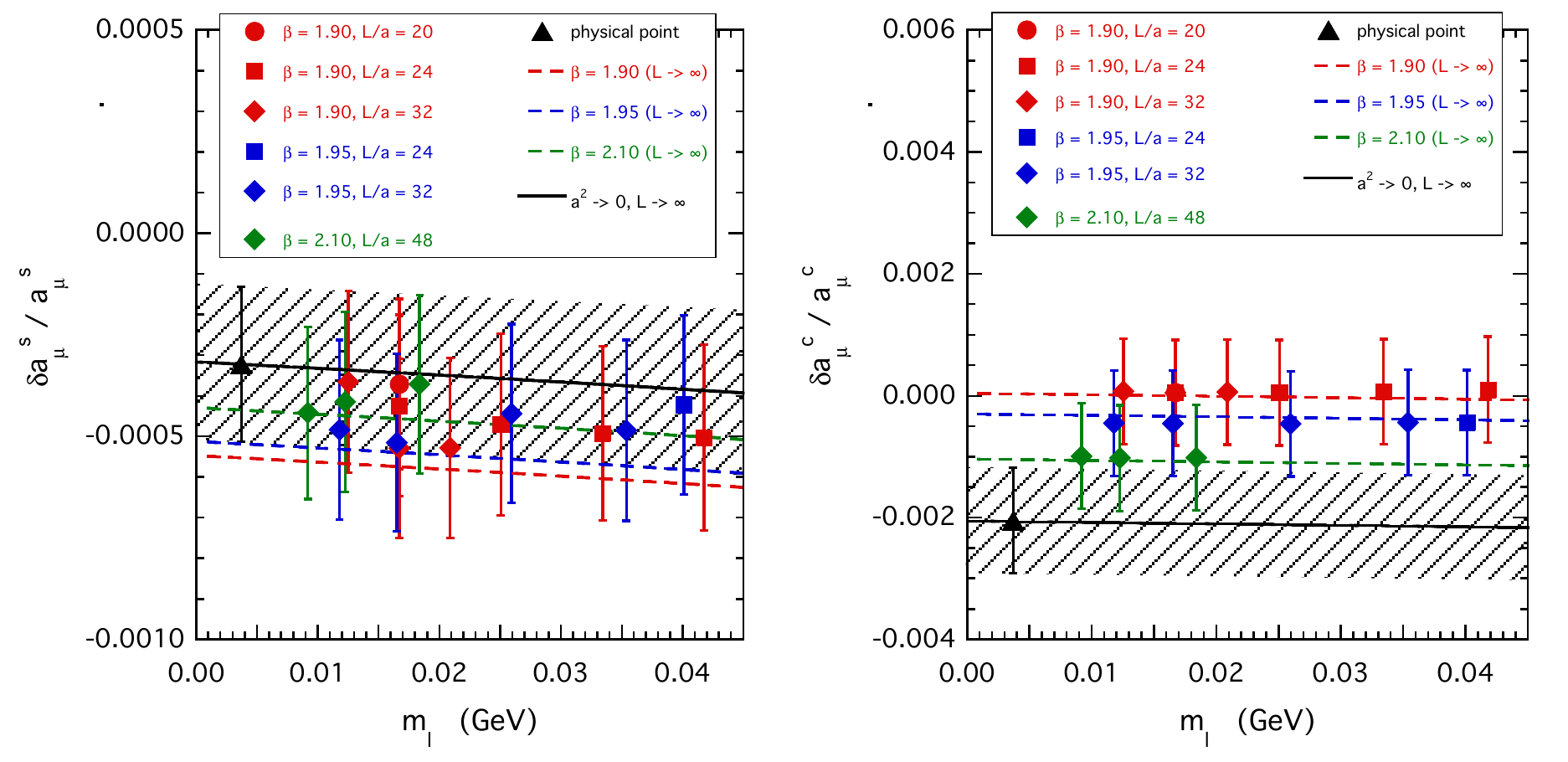}}}
\caption{\it \small Results for the strange (left panel) and charm (right panel) contributions to $\delta a_\mu^{had} / a_\mu^{had}$, obtained for the ETMC gauge ensembles of Table \ref{tab:simudetails}. The dashed lines correspond to the linear fit (\ref{eq:fit_damuq}) including the discretization term in the infinite volume limit. The solid lines correspond to the linear fit in the continuum and infinite volume limits, while the shaded areas identify the corresponding uncertainty at the level of one standard deviation. The triangles are the results of the extrapolation at the physical pion mass and in the continuum and infinite volume limits. The data are evaluated without adopting the ELM procedure (\ref{eq:HLtrick}).}
\label{fig:ratioq_fit}
\end{figure}

The precision of the lattice data can be drastically improved by forming the ratio of the e.m.~correction over the lowest-order term. 
Therefore, in what follows we perform our analysis of the ratio $\delta a_\mu^{had} / a_\mu^{had}$, which is shown in Fig.~\ref{fig:ratioq_fit}.
We have checked that in the case of the e.m.~corrections the use of the ELM procedure (\ref{eq:HLtrick}) does not improve the precision of the lattice data.

It can be seen from Fig.~\ref{fig:ratioq_fit} that the dependence on the light-quark mass $m_\ell$ is quite mild, being driven only by sea quarks, and that the uncertainties of the data are dominated by the error on the RC $\delta Z_A$, which has been taken to be the same for all the gauge ensembles used in this work (see Appendix \ref{sec:ZA}).

The FSEs are visible only in the case of the strange quark.
A theoretical calculation of FSEs for $\delta a_\mu^{had}$ is not yet available.
According to the general findings of Ref.~\cite{Lubicz:2016xro} the universal FSEs are expected to vanish, since they depend on the global charge of the meson states appearing in the spectral decomposition of the vector correlator $\delta V(t)$.
Moreover, the structure-dependent (SD) FSEs are expected to start at order ${\cal{O}}(1/L^2)$.
According to the effective field theory approach of Ref.~\cite{Davoudi:2014qua}, one might argue that in the case of mesons with vanishing charge radius (as the ones appearing in the correlator $\delta V(t)$) the SD FSEs may start at order ${\cal{O}}(1/L^3)$.
Therefore we adopt the following simple fitting function
 \be
     \frac{\delta a_\mu^{s,c}}{a_\mu^{s,c}} = \delta A_0^{s,c} + \delta A_1^{s,c} m_\ell + \delta D^{s,c} a^2 + \delta F^{s,c} \frac{1}{L^n} 
     \label{eq:fit_damuq}
 \ee
where the power $n$ can be put equal to $n = 2$ or $n = 3$.
In fitting our data we do not observe sensitivity to the above choices of the power $n$ within the statistical uncertainties.

At the physical pion mass and in the continuum and infinite volume limits we get
 \bea
     \frac{\delta a_\mu^{s, phys}}{a_\mu^{s, phys}} & = & -0.000332 ~ (46)_{stat+fit} ~ (6)_{input} ~ (8)_{FSE} ~ (4)_{chir} ~ (2)_{disc} ~ (208)_{Z_A} ~ , \nonumber \\
                                                                               & = & -0.000332 ~ (46)_{stat+fit} ~ (208)_{syst} ~ , \nonumber \\
                                                                               & = & -0.000332 ~ (213) ~ ,   
     \label{eq:damus_phys}
 \eea
and
 \bea
     \frac{\delta a_\mu^{c, phys}}{a_\mu^{c, phys}} & = & -0.00205 ~ (12)_{stat+fit} ~ (1)_{input} ~ (1)_{FSE} ~ (1)_{chir} ~ (1)_{disc} ~ (85)_{Z_A} ~ , \nonumber \\
                                                                               & = & -0.00205 ~ (12)_{stat+fit} ~ (85)_{syst} ~ , \nonumber \\
                                                                               & = & -0.00205 ~ (86) ~ ,
     \label{eq:damuc_phys}  
 \eea
where
\begin{itemize}
\item $()_{stat+fit}$ indicates the uncertainty induced by both the statistical errors and the fitting procedure itself;
\item $()_{input}$ is the error coming from the uncertainties of the input parameters of the eight branches of the quark mass analysis of Ref.~\cite{Carrasco:2014cwa};
\item $()_{disc}$ is the uncertainty due to both discretization effects and scale setting, estimated by comparing the results obtained with and without the ELM procedure (\ref{eq:HLtrick});
\item $()_{FSE}$ is the error coming from including ($\delta F^s \neq 0$) or excluding ($\delta F^s = 0$) the FSE correction. When FSEs are not included, all the gauge ensembles with $L / a = 24$ are also not included;
\item $()_{chir}$ is the error coming from including ($\delta A_1^s \neq 0$) or excluding ($\delta A_1^s = 0$) the linear term in the light-quark mass.
\item $()_{Z_A}$ is the error generated by the uncertainty on the RC $Z_A^{fact}$ (see Eq.~(\ref{eq:deltaZA})), which turns out to be by far the dominant source of uncertainty.
\end{itemize}

Using the lowest-order results (\ref{eq:amus_phys}-\ref{eq:amuc_phys}) we obtain 
 \bea
     \label{eq:damus_phys}
     \delta a_\mu^{s, phys} & = & - 0.018 ~ (11) \cdot 10^{-10} ~ , \\
     \label{eq:damuc_phys}
     \delta a_\mu^{c, phys} & = & - 0.030 ~ (13) \cdot 10^{-10} ~ .    
 \eea

Thus, the e.m.~corrections to $\delta a_\mu^s$ and $\delta a_\mu^c$ turn out to be negligible with respect to the current uncertainties of the lowest-order terms.

\section{Conclusions}
\label{sec:conclusions}

We have presented a lattice calculation of the HVP contribution of strange and charm quarks to the anomalous magnetic moment of the muon at orders ${\cal{O}}(\alpha_{em}^2)$ and ${\cal{O}}(\alpha_{em}^3)$ in the e.m.~coupling. 

We have employed the gauge configurations generated by the European Twisted Mass Collaboration with $N_f = 2+1+1$ dynamical quarks at three values of the lattice spacing ($a \simeq 0.062 - 0.089$ fm) with pion masses in the range $M_\pi \simeq 210 - 450$ MeV and with strange and charm quark masses tuned at their physical values.

In this work we have taken into account only connected diagrams, in which each quark flavor contributes separately, and a direct summation of the relevant correlators over the Euclidean time distances has been performed, adopting the local lattice version of the e.m.~current operator.

As for the calculation of the e.m.~corrections in the strange and charm sectors, we have adopted the RM123 approach of Ref.~\cite{deDivitiis:2013xla}, based on the expansion of the lattice path-integral in powers of the {\it small} e.m.~coupling, namely $\alpha_{em} \approx 1 \%$.

After extrapolation to the physical pion mass and to the continuum limit our results for $a_\mu^{had}$ are for the lowest-order contributions
 \bea
        a_\mu^s(\alpha_{em}^2) & = & (53.1 \pm 2.5) \cdot 10^{-10} ~ , \\
        a_\mu^c(\alpha_{em}^2) & = & (14.75 \pm 0.56) \cdot 10^{-10}
 \eea
and for the e.m.~corrections
 \bea
        a_\mu^s(\alpha_{em}^3) & = & (-0.018 \pm 0.011) \cdot 10^{-10} ~ , \\
        a_\mu^c(\alpha_{em}^3) & = & (-0.030 \pm 0.013) \cdot 10^{-10} ~ ,      
 \eea 
which show that the latter ones are negligible with respect to the present uncertainties of the lowest-order terms.
We stress that the current uncertainties on the e.m.~corrections $\delta a_\mu^s$ and $\delta a_\mu^c$ are of the order of $\sim 60 \%$ and $\sim 40 \%$, since they are dominated by the uncertainty on the RC $Z_A$ of the local vector current, which has been estimated through the axial Ward-Takahashi identity (WTI) derived in the presence of QED effects (see Appendix \ref{sec:ZA}).
A dedicated study aimed at the determination of the RCs of bilinear operators in presence of QED employing non-perturbative renormalization schemes, like the RI-MOM one, is expected to improve significantly the precision of the calculation of the e.m.~corrections and isospin-breaking effects on $a_\mu^{had}$.

Our findings demonstrate that the expansion method of Ref.~\cite{deDivitiis:2013xla}, which has been already applied successfully to the calculation of e.m.~corrections to meson masses~\cite{deDivitiis:2013xla,Giusti:2017dmp} and to the leptonic decays of pions and kaons \cite{Lubicz:2016mpj,Tantalo:2016vxk}, works as well also in the case of the HVP contribution to the muon anomalous magnetic moment.
The application of the approach presented in this work to the case of the $u$- and $d$-quark contributions is ongoing.

\section*{Acknowledgements}
We warmly thank our colleagues of the ETM collaboration, in particular R.~Frezzotti, K.~Jansen, M.~Petschlies, G.C.~Rossi, N. Tantalo and C.~Tarantino, for many fruitful discussions and comments. 
We gratefully acknowledge the CPU time provided by PRACE under the project Pra10-2693 {\em ``QED corrections to meson decay rates in Lattice QCD''} and by CINECA under the specific initiative INFN-LQCD123 on the BG/Q system Fermi at CINECA (Italy).
V.L., G.M., S.S.~thank MIUR (Italy) for partial support under the contract PRIN 2015. 
G.M.~also acknowledges partial support from ERC Ideas Advanced Grant n. 267985 ``DaMeSyFla''.

\appendix

\section{Non-perturbative estimate of the RCs $Z_V^{(fact)}$ and $Z_A^{(fact)}$}
\label{sec:ZA}

\subsection{Axial Ward-Takahashi identity in the presence of electromagnetism}

For an isospin doublet $\psi \equiv \left( \psi_1, \psi_2 \right)$ of mass-degenerate quarks the twisted-mass (TM) action including QED is given in the physical basis at maximal twist by \cite{Frezzotti:2000nk,Frezzotti:2003ni,Frezzotti:2004wz}
 \be
    {\cal{S}} = a^4 \sum_x \overline{\psi}(x) \left\{ M(x, y) - \frac{1}{2a} \sum_\mu \left[F_\mu(x, y) + B_\mu(x, y) \right] \right\} \psi(y) ~ ,
 \ee
where
 \bea
    M(x, y) & = & \left[ m + i \left( 4\frac{r}{a} - m^{crit} \right) \gamma_5 \tau_3 \right] \delta(x, y) ~ , \\[2mm]
    F_\mu(x, y) & = & \left( ir \gamma_5 \tau_3 - \gamma_\mu \right) U_\mu(x) E_\mu(x) \delta(x + a \hat{\mu}, y) ~ , \\[2mm]
    B_\mu(x, y) & = & \left( ir \gamma_5 \tau_3 + \gamma_\mu \right) U_\mu^\dagger(y) E_\mu^\dagger(y) \delta(x - a \hat{\mu}, y) ~ ,
 \eea
with $E_{\mu}(x) = e^{ieQ {\cal{A}}_\mu(x)}$ being the QED link, ${\cal{A}}_{\mu}\left(x\right)$ the photon field, $m$ the twisted bare quark mass (in QCD+QED), $m^{crit} $ the critical mass (in QCD+QED) and $Q \equiv {\rm diag}\left\{ q_1, q_2 \right\}$.
Performing the local non-singlet axial rotation
 \bea
     \psi(x) & \to & \left[1 + i \alpha(x)\tau^+ \gamma_5 \right] \psi(x) \nonumber \\[2mm]
    \overline{\psi}(x) & \to & \overline{\psi}(x) \left[1 + i \alpha(x) \tau^+ \gamma_5 \right] \nonumber
\eea
with $\tau^+ \equiv \tau_1 + i \tau_2$, one has
 \be
    \label{eq:WITM}
    \frac{\delta S}{\delta\left[ i \alpha(x) \right]} = - \partial_\mu A_\mu(x) + 2m ~ \overline{\psi}(x) \gamma_5 \tau^+ \psi(x) + X(x) = 0 ~ ,
 \ee
where $\partial_\mu$ is the backward derivative in the $\mu$ direction and
 \bea
    \label{eq:ATM}
    A_\mu(x) & = & \frac{1}{2} \left[ \overline{\psi}(x) \gamma_\mu \gamma_5 \tau^+ U_\mu(x) E_\mu(x) \psi(x + a\hat{\mu}) +\rm{h.c.} \right] \\
    \label{eq:X}
    X(x) & = & -\frac{1}{2a} \sum_\mu \overline{\psi}(x - a\hat{\mu}) \gamma_\mu \gamma_5 U_\mu(x - a\hat{\mu}) \left[\tau^+, ~ E_\mu(x - a\hat{\mu}) \right] \psi(x) + 
                     \rm{h.c.} \nonumber \\
            & - & \frac{ir}{2a}\sum_{\mu}\overline{\psi}\left(x\right) \tau^+ \tau^3 U_{\mu}\left(x\right)E_{\mu}\left(x\right)\psi\left(x+a\hat{\mu}\right) + \rm{h.c.} \nonumber \\
            & - & \frac{ir}{2a}\sum_{\mu}\overline{\psi}\left(x-a\hat{\mu}\right) \tau^3 U_{\mu}\left(x-a\hat{\mu}\right)E_{\mu}\left(x-a\hat{\mu}\right)\tau^+ \psi\left(x\right) + 
                     \rm{h.c.}
 \eea

\noindent We now choose that the charges of the two quarks are the same, i.e.~$q_1 = q_2 = q$.
This implies that the isospin rotation $\tau^+$ commutes with the QED link $E_\mu(x)$.
Consequently, the first line in Eq.~(\ref{eq:X}) vanishes, while the second and third lines can be written as a backward derivative.
Thus, Eq.~(\ref{eq:WITM}) becomes
 \be
    \label{eq:WITM_final}
    \partial_\mu \overline{A}_\mu^{TM}(x) = 2m ~ \overline{\psi}(x) \gamma_5 \tau^+ \psi(x) ~ ,
 \ee
where $ \overline{A}_\mu^{TM}(x)$ is the 1-point split TM axial current
 \bea
    \overline{A}_\mu^{TM}(x) & = & \frac{1}{2} \left[ \overline{\psi}(x) \gamma_\mu \gamma_5 \tau^+ U_\mu(x) E_\mu(x) \psi(x + a \hat{\mu}) \right. \nonumber \\
                                             & - & \left. i r \overline{\psi}(x) \tau^+ \tau^3 U_\mu(x) E_\mu(x) \psi(x + a \hat{\mu}) + \rm{h.c.} \right] ~ , 
    \label{eq:ATM_conserved}
 \eea
which is conserved in the chiral limit and therefore it does not require any renormalization constant.

As is well known, the local current 
 \be
    \label{eq:ATM_local}
    A_\mu^{TM}(x) \equiv \overline{\psi}(x) \gamma_\mu \gamma_5 \tau^+ \psi(x) 
 \ee
requires a multiplicative renormalization, given by the RC $Z_V$~\cite{Frezzotti:2003ni}, in order to match the 1-point split TM axial current (\ref{eq:ATM_conserved}) in the continuum limit.
Thus, provided the quark charges are the same, the local version of the TM axial Ward-Takahashi identity holds as well also in the presence of electromagnetism, viz. 
 \be
    Z_V ~ \partial_\mu A_\mu^{TM}(x) = 2m ~ P_5^{TM}(x) + {\cal{O}}(a^2) ~ ,
    \label{eq:Ax_WI}
 \ee
where $P_5^{TM}(x)$ is the bare pseudoscalar density operator $P_5^{TM}(x) \equiv  \overline{\psi}(x) \gamma_5 \tau^+ \psi(x)$.

\subsection{Determination of the RC $Z_V$}

Let's consider a pseudoscalar (PS) meson composed by the two mass- and charge-degenerate TM quarks $\left( \psi_1, \psi_2 \right)$. 
Introducing the 2-point correlators
 \bea
    \label{eq:P5P5}
    C_{P_5^{TM} P_5^{TM}}(t) & = & \sum_{\vec{x}} \left\langle P_5^{TM}(\vec{x}, t) P_5^{TM \dagger}(0) \right\rangle ~ , \\
    \label{eq:A0P5}
    C_{A_0^{TM} P_5^{TM}}(t) & = & \sum_{\vec{x}} \left\langle A_{0}^{TM}(\vec{x}, t) P_5^{TM \dagger}(0) \right\rangle ~ ,
 \eea
Eq.~(\ref{eq:Ax_WI}) implies
 \be
     Z_V ~ \partial_t C_{A_0^{TM} P_5^{TM}}(t) = 2m ~ C_{A_0^{TM} P_5^{TM}}(t) + {\cal{O}}(a^2) ~ .
 \ee
At large Euclidean time distances $t >> a$ one has
 \bea
    \label{eq:P5P5_larget}
    C_{P_5^{TM} P_5^{TM}}(t) & ~_{\overrightarrow{t >> a}} ~ & \frac{| \langle 0 | P_5^{TM} |PS \rangle |^2}{2M_{PS}^{TM}} 
         \left[ e^{- M_{PS}^{TM} t} + e^{- M_{PS}^{TM} (T - t)} \right] ~ , \\
    \label{eq:A0P5_larget}
    C_{A_0^{TM} P_5^{TM}}(t) & ~_{\overrightarrow{t >> a}} ~ & \frac{\langle 0 | A_0^{TM} |PS \rangle \langle 0 | P_5^{TM} |PS \rangle^*}{2M_{PS}^{TM}} 
         \left[ e^{- M_{PS}^{TM} t} - e^{- M_{PS}^{TM} (T - t)} \right] ~ ,
 \eea
and therefore the renormalization constant $Z_V$ can be determined in terms of the matrix elements $\langle 0 | P_5^{TM} |PS \rangle$ and $\langle 0 | A_0^{TM} |PS \rangle$ as
 \be
    \label{eq:ZV}
    Z_V = \frac{2 m}{M_{PS}^{TM}} \frac{\langle 0 | P_5^{TM} |PS \rangle}{\langle 0 | A_0^{TM} |PS \rangle} ~ .
 \ee

For a generic quantity $O$ we consider the following expansion in $\alpha_{em}$
 \be
    O = O^{(0)} + \delta O + {\cal{O}}(\alpha_{em}^2) ~ ,
 \ee
where $O^{(0)}$ and $\delta O$ indicates the quantity in absence of QED and at ${\cal{O}}(\alpha_{em})$, respectively .
Thus, from Eq.~(\ref{eq:ZV}) one gets
\be
\frac{\delta Z_V}{Z_V^{(0)}} = \frac{\delta m}{m^{(0)}} + \frac{\delta \langle 0 | P_5^{TM} |PS \rangle}{\langle 0 | P_5^{TM} |PS^{(0)} \rangle} - 
    \frac{\delta M_{PS}^{TM}}{M_{PS^{(0)}}^{TM}} - \frac{\delta \langle 0 | A_0^{TM} |PS \rangle}{\langle 0 | A_0^{TM} |PS^{(0)} \rangle} ~ ,
\label{eq:ZV_prime}
\ee
where $\delta m$ arises from the ${\cal{O}}(\alpha_{em})$ contribution to the bare quark mass given by Eq.~(\ref{eq:Zf}), viz.
\be
     \delta m = m^{(0)} ~ \frac{\alpha_{em} q^2}{4 \pi} \left[ 6 \mbox{log}(a \mu) - 22.5954 \right] ~ .
\ee  

The quantities $\delta \langle 0 | P_5^{TM} |PS \rangle$, $\delta M_{PS}^{TM}$ and $\delta \langle 0 | A_0^{TM} |PS \rangle$ can be extracted from the contributions at ${\cal{O}}(\alpha_{em})$ to the 2-point correlators (\ref{eq:P5P5}-\ref{eq:A0P5}).
Putting $X = \{P_{5}^{TM}, A_{0}^{TM} \}$ one has
 \be
    \delta C_{XP_5^{TM}}(t) = \delta C_{XP_5^{TM}}^{exch}(t) + \delta C_{XP_5^{TM}}^{self}(t) +  \delta C_{XP_5^{TM}}^{tad}(t) + 
                                               \delta C_{X P_5^{TM}}^{PS}(t) + \delta C_{X P_5}^S(t) ~ ,
    \label{eq:ratio}
 \ee
where the superscripts refer to the exchange, self-energy, tadpole, PS and scalar insertions, introduced already in Eqs.~(\ref{eq:deltaV}-\ref{eq:deltaV_S}) and depicted in Fig.~\ref{fig:diagrams}.

The ratio of Eq.~(\ref{eq:ratio}) with the lowest-order correlator $C_{XP_{5}^{TM}}^{(0)}(t)$ behaves at large Euclidean time distances (up to around-the-world effect) as
\be
\frac{\delta C_{X P_5^{TM}}(t)}{C_{XP_{5}^{TM}}^{(0)}(t)} ~_{\overrightarrow{t >> a}} ~ 
    \frac{\delta \left[ \langle 0 | P_5^{TM} |PS \rangle \langle 0 | X |PS \rangle \right]}{\langle 0 | P_5^{TM} |PS^{(0)} \rangle \langle 0 | X |PS^{(0)} \rangle} - 
    \delta M_{PS}^{TM} \cdot t 
\label{eq:corr_2pts_ins}
\ee
from which all the ingredients entering Eq.~(\ref{eq:ZV_prime}) can be calculated.

\subsection{Determination of the RCs $Z_A$ and $Z_P / Z_S$}

We now consider an isospin doublet $\psi^{OS} = \left( \psi_1^{OS}, \psi_2^{OS} \right)$ of mass- and charge-degenerate quark fields regularized using the Osterwalder-Seiler (OS) prescription \cite{Osterwalder:1977pc}, i.e.~the same value of the Wilson $r$-parameter is assumed for the two quarks.
 At maximal twist the local axial current 
 \be
    \label{eq:AOS_local}
    A_\mu^{OS}(x) \equiv \overline{\psi}^{OS}(x) \gamma_\mu \gamma_5 \tau^+ \psi^{OS}(x) 
 \ee
requires a multiplicative RC given by $Z_A$~\cite{Frezzotti:2003ni}.
Once renormalized the matrix elements of the TM (\ref{eq:ATM_local}) and OS (\ref{eq:AOS_local}) axial currents can differ only by discretization effects and therefore one has 
\be
    Z_A \langle 0 | A_0^{OS} |PS \rangle = Z_V \langle 0 | A_0^{TM} |PS \rangle + {\cal{O}}(a^2) ~ .
\ee
This implies 
\bea
\frac{\delta Z_A}{Z_A^{(0)}} & = & \frac{\delta Z_V}{Z_V^{(0)}} + \frac{\delta \langle 0 | A_0^{TM} |PS \rangle}{\langle 0 | A_0^{TM} |PS^{(0)} \rangle} - 
    \frac{\delta \langle 0 | A_0^{OS} |PS \rangle}{\langle 0 | A_0^{OS} |PS^{(0)} \rangle} \nonumber \\ 
    & = & \frac{\delta m}{m^{(0)}} + \frac{\delta \langle 0 | P_5^{TM} |PS \rangle}{\langle 0 | P_5^{TM} |PS^{(0)} \rangle} - 
    \frac{\delta M_{PS}^{TM}}{M_{PS^{(0)}}^{TM}} - \frac{\delta \langle 0 | A_0^{OS} |PS \rangle}{\langle 0 | A_0^{OS} |PS^{(0)} \rangle} ~ ,
\label{eq:ZA_prime}
\eea
where $\delta \langle 0 | A_0^{OS} |PS \rangle / \langle 0 | A_0^{OS} |PS^{(0)} \rangle$ can be determined from the relevant axial correlators computed in the OS regularization.

Similarly, the ratio $Z_P / Z_S$ of the RCs of the pseudoscalar and scalar densities can be determined by using the relation
\be
    Z_S \langle 0 | P_5^{OS} |PS \rangle = Z_P \langle 0 | P_5^{TM} |PS \rangle + {\cal{O}}(a^2) ~ .
\ee
As for the e.m.~corrections $\delta Z_P$ and $\delta Z_S$ one has
\be
    \label{eq:delta_ZPS}
    \frac{\delta Z_P}{Z_P^{(0)}} - \frac{\delta Z_S}{Z_S^{(0)}} = \frac{\delta \langle 0 | P_5^{OS} |PS \rangle}{\langle 0 | P_5^{OS} |PS^{(0)} \rangle} - 
                                                    \frac{\delta \langle 0 | P_5^{TM} |PS \rangle}{\langle 0 | P_5^{TM} |PS^{(0)} \rangle} ~ ,
\ee
which, however, does not allow to determine separately the two corrections $\delta Z_P$ and $\delta Z_S$.
For this reason we will not investigate Eq.~(\ref{eq:delta_ZPS}) numerically.

\subsection{Numerical results}

In order to get a first non-perturbative estimate of the RCS $Z_V$ and $Z_A$ in QCD+QED we have calculated the r.h.s.~of Eqs.~(\ref{eq:ZV_prime}) and (\ref{eq:ZA_prime}) using for the (bare) quark mass $m$ the values of the strange quark masses reported in Table \ref{tab:simudetails} for the three values of the inverse lattice coupling $\beta$ of the ETMC ensembles.
As in Section~\ref{sec:deltas&c}, we introduce the correction factors $Z_V^{(fact)}$ and $Z_A^{(fact)}$ to the ``naive factorization'' approximation by defining
 \be
     \delta Z_V = Z_V^{(0)} \cdot Z_V^{(em)} \cdot Z_V^{(fact)} ~ , \qquad \delta Z_A = Z_A^{(0)} \cdot Z_A^{(em)} \cdot Z_A^{(fact)} ~ ,
     \label{eq:Zfact}
 \ee
where $Z_{V(A)}^{(em)}$ is the one-loop perturbative estimate of the QED effect at order ${\cal{O}}(\alpha_s^0)$ in the strong coupling.
Using for the latter ones the perturbative findings $Z_V^{(em)} = - 20.6178 ~ \alpha_{em} q^2 / (4 \pi)$ and $Z_A^{(em)} = - 15.7963 ~ \alpha_{em} q^2 / (4 \pi)$ from Refs.~\cite{Martinelli:1982mw,Aoki:1998ar}, our results for $Z_V^{(fact)}$ and $Z_A^{(fact)}$ are collected in Table \ref{tab:fact}.
\begin{table}[hbt!]
\begin{center}
\begin{tabular}{||c||c|c||}
\hline 
$\beta$ & $Z_V^{(fact)}$ & $Z_A^{(fact)}$ \\
\hline \hline 
1.90 & 1.027 (5) & 0.85 (5) \\
\hline 
1.95 & 1.033 (4) & 0.93 (5) \\
\hline 
2.10 & 1.034 (3) & 0.87 (6) \\
\hline
\end{tabular}
\end{center}
\caption{\it \small Results for $Z_V^{(fact)}$ and $Z_A^{(fact)}$ (see Eq.~(\ref{eq:Zfact})) obtained at the three values of the inverse bare lattice coupling $\beta$ corresponding to the gauge ensembles of Table \ref{tab:simudetails}.}
\label{tab:fact}
\end{table}
It can be seen that the dependence on the lattice spacing is quite mild within the uncertainties.
The averages of the results of Table \ref{tab:fact} (according to Eq.~(28) of Ref.~\cite{Carrasco:2014cwa}) are: $Z_V^{(fact)} = 1.031 \pm 0.005$ and $Z_A^{(fact)} = 0.88 \pm 0.06$.
Given the exploratory nature of the present non-perturbative determination of QCD+QED renormalization constants, we prefer to quote as our estimates for $Z_V^{(fact)}$ and $Z_A^{(fact)}$ the values
\be
     Z_V^{(fact)} = 1.03 \pm 0.01 ~ , \qquad Z_A^{(fact)} = 0.9 \pm 0.1 ~ ,
\ee
which cover the spread of the results given in Table~\ref{tab:fact}.

\end{document}